\documentclass{aastex61}
\usepackage{graphicx}
\usepackage{longtable}
\usepackage{multirow}
\usepackage{amsmath}
\usepackage{url}
\usepackage{amsmath,bm}
\usepackage{lipsum}
\usepackage{rotating}

\date{\today}

\begin{document}

\title{Dark matter deficient galaxies in the Illustris flat-$\Lambda$CDM model structure formation simulation}

\author{Hai Yu}
\altaffiliation{yuhai@smail.nju.edu.cn}
\affiliation{School of Astronomy and Space Science, Nanjing University, Nanjing 210093, China}
\affiliation{Department of Physics, Kansas State University, 116 Cardwell Hall, Manhattan, KS 66506, USA}

\author{Bharat Ratra}
\altaffiliation{ratra@phys.ksu.edu}
\affiliation{Department of Physics, Kansas State University, 116 Cardwell Hall, Manhattan, KS 66506, USA}

\author{Fa-Yin Wang}
\altaffiliation{fayinwang@nju.edu.cn}
\affiliation{School of Astronomy and Space Science, Nanjing University, Nanjing 210093, China}
\affiliation{Key Laboratory of Modern Astronomy and Astrophysics (Nanjing University), Ministry of Education, Nanjing 210093, China}

\begin{abstract}
Surveying dark matter deficient galaxies (those with dark matter mass to 
stellar mass ratio $M_{\rm dm}/M_{\rm star}<1$) in the Illustris simulation 
of structure formation in the flat-$\Lambda$CDM cosmogony, we find 
$M_{\rm star} \approx 2 \times 10^8\, M_\sun$ galaxies that have properties 
similar to those ascribed by \citet{vanDokkumetal2018a} to the ultra-diffuse
galaxy NGC1052-DF2. The Illustris simulation also contains more luminous 
dark matter deficient galaxies.
Illustris galaxy subhalo 476171 is a particularly interesting outlier, a 
massive and very compact galaxy with 
$M_{\rm star} \approx 9 \times 10^{10}\, M_\sun$ and 
$M_{\rm dm}/M_{\rm star} \approx 0.1$ and a half-stellar-mass radius of 
$\approx 2$ kpc. If the Illustris simulation and 
the $\Lambda$CDM model are accurate, there are a significant number of 
dark matter deficient galaxies, including massive luminous compact ones. 
It will be interesting to observationally discover these galaxies, and to 
also more clearly understand how they formed, as they are likely to provide 
new insight into and constraints on models of structure formation and the 
nature of dark matter. 
\end{abstract}


\section{Introduction}\label{sec:introduction}

In the standard $\Lambda$CDM model \citep{Peebles1984} the cosmological 
constant $\Lambda$ powers the currently accelerating cosmological
expansion and cold dark matter (CDM) is now the second biggest 
contributor to the cosmological energy budget. Earlier, when nonrelativistic 
CDM and baryonic matter dominated, the cosmological expansion decelerated. The 
standard spatially-flat $\Lambda$CDM model is consistent with many 
observational constraints when the current $\Lambda$, CDM, and baryonic 
density parameters are at or near $\Omega_\Lambda = 0.70$, $\Omega_{\rm c} = 
0.25$, and $\Omega_{\rm b} = 0.05$ \citep{ParkRatra2018b}.\footnote{For 
reviews of the standard model see \citet{RatraVogeley2008}, 
\citet{Martin2012}, and \citet{Lukovicetal2018}.}

These observations include cosmic microwave background (CMB) anisotropies 
\citep{Hinshawetal2013, PlanckCollaboration2016}, baryon acoustic oscillation 
distances \citep{Alametal2017, Ryanetal2018}, supernova Type Ia apparent 
magnitudes \citep{Scolnicetal2017}, and Hubble parameters 
\citep{Farooqetal2017, Yuetal2018}. There is significant observational 
evidence for CDM, as well as for $\Lambda$ (or a dark energy that behaves 
almost like $\Lambda$), 
and introducing these fairly astonishing hypothetical substances appears 
to be the most reasonable way to make sense of the observations. For instance, 
Hubble parameter observations span a wide redshift range, almost to $z = 2.4$,
and show evidence for both a present epoch dark energy powered accelerating 
cosmological expansion as well as an earlier CDM and baryonic matter driven 
decelerated cosmological expansion \citep{FarooqRatra2013, Farooqetal2013, Morescoetal2016, Farooqetal2017, Yuetal2018, Jesusetal2018, Haridasuetal2018}. It 
is also widely accepted that CDM is necessary for the formation of observed 
structure in the cosmological matter and radiation fields.  

In the standard CDM structure formation model \citep{Peebles1982}, quantum 
zero-point fluctuations of the inflaton scalar field during 
inflation \citep{Hawking1982, Starobinsky1982, GuthPi1982, Fischleretal1985} 
seeded spatial inhomogeneities that then grew under gravitational 
instability to create the observed structures in the cosmological matter 
and radiation fields. Initially inhomogeneities developed and grew in the CDM; 
as the CMB cooled and decoupled from the baryons, baryonic matter was 
gravitationally attracted to and fell deeper into the CDM halo inhomogeneity 
gravitational potential wells.  While the initial gathering of CDM 
under gravity is a relatively simple process, when the baryons start to play 
a significant role the structure formation problem becomes much less 
tractable because the physics is more familiar, more complex, and more 
difficult to model quantitatively.       

In the standard $\Lambda$CDM model it is not inconceivable that after a 
collection of baryonic matter formed stars and became a ``galaxy'', this 
galaxy might be ejected from the CDM halo in which it was formed and could 
find itself in a region with a lower CDM density.\footnote{It is highly 
unlikely that baryonic structure, a ``galaxy'', could form in very low CDM 
density regions.}
It is not known how probable such an outcome is. The \citet{vanDokkumetal2018a} 
observations and argument that the ultra-diffuse dwarf galaxy NGC1052-DF2 
has a stellar mass of about $2 \times 10^8\, M_\sun$ and a total mass of less 
than $3.4 \times 10^8\, M_\sun$ in a radius of 7.6 kpc and so indicates a much 
smaller dark matter mass to stellar mass ratio, $M_{\rm dm}/M_{\rm star}$, than 
that of a typical galaxy with this stellar mass,\footnote{We note that there 
has been some discussion of these observations and the 
\citet{vanDokkumetal2018a} interpretation of them \citep{vanDokkumetal2018b, Martinetal2018, Laporteetal2018, Famaeyetal2018, Scarpaetal2018, Nusser2018, Trujilloetal2018, vanDokkumetal2018c, Wassermanetal2018, BlakesleeCantiello2018}.} 
motivated us to attempt to determine the probability of such an outcome.

We study the issue of dark matter deficient galaxies by looking at galaxy 
population statistics of the Illustris simulation 
\citep[www.illustris-project.org,][]{Vogelsbergeretal2014a} of structure 
formation in the flat-$\Lambda$CDM cosmogony. We find that dark matter 
deficient galaxies, with $M_{\rm star} \approx 2 \times 10^8\, M_\sun$ and 
$M_{\rm dm}/M_{\rm star}<1$ \citep[like NGC1052-DF2,][]{vanDokkumetal2018a}, 
are not uncommon. We also find a significant number of more luminous and 
more massive dark matter deficient galaxies.

\section{Dark matter deficient galaxy statistics and examples}\label{sec:simulation}

Illustris \citep[www.illustris-project.org,][]{Vogelsbergeretal2013, Vogelsbergeretal2014a, Vogelsbergeretal2014b, Geneletal2014, Sijackietal2015}
is one of the largest hydrodynamical cosmological simulations. It can
help us understand how structure in the universe evolves with time, and 
in particular how the dark matter and stellar mass distributions evolve. The 
simulation assumes a tilted spatially-flat $\Lambda$CDM cosmogony with 
chosen parameter values in reasonable accord with current cosmological 
measurements \citep{ParkRatra2018b}.\footnote{Current cosmological data are 
also not inconsistent with mildly closed spatial hypersurfaces 
\citep{Oobaetal2018a, ParkRatra2018a, ParkRatra2018b, ParkRatra2018d}, 
with mild dark energy dynamics \citep{Oobaetal2018c, ParkRatra2018b, ParkRatra2018c, ParkRatra2018d}, and with 
nonflat dynamical dark energy models \citep{Oobaetal2017, Oobaetal2018b, ParkRatra2018b, ParkRatra2018c, ParkRatra2018d}. 
While reionization is 
remarkably different in the closed model compared to the standard flat 
case \citep{Mitraetal2018}, it is likely that structure formation will 
be less affected by observationally-consistent values of nonzero spatial 
curvature or dark energy dynamics.} The fiducial parameter values
chosen for the simulation are $(\Omega_{\rm m},\Omega_\Lambda,\Omega_{\rm b},\sigma_8,n_{\rm s},h)=(0.2726,0.7274,0.0456,0.809,0.963,0.704)$ coming from 
the final WMAP analysis \citep{Hinshawetal2013}.\footnote{Here $\Omega_{\rm m}$
is the nonrelativistic matter density parameter, $\sigma_8$ is the rms
fractional energy density inhomogeneity averaged over 8$h^{-1}$ Mpc spheres,
$n_{\rm s}$ is the spectral index of the primordial energy density perturbation 
power spectrum (which is assumed to be a power law in wavenumber), and $h$ is
the Hubble constant in units of 100 km s$^{-1}$ Mpc$^{-1}$.}  
The Illustris-1 simulation contains 1820$^3$ dark matter particles, 
1820$^3$ gas particles, and 1820$^3$ tracer particles in a comoving box 
of size (106.5\, Mpc)$^3$. The mass of each dark matter particle is 
$6.26\times10^6\,M_\odot$ and for baryonic particles the mass is 
$1.26\times10^6\,M_\odot$. The simulation not only accounts for gravity and 
hydrodynamics but also includes effects of star formation and evolution, 
gas cooling, black holes and supermassive black hole feedback, as well as 
other relevant phenomena, and so should allow us to have a fuller 
understanding of the effects of various physical processes on the formation 
of large-scale structure. In this work we make use of the Illustris-1 
simulation.

At redshift $z = 0$ there are 7,713,601 friends-of-friends (FoF) groups (with 
more than 32 dark matter particles) and 4,366,546 individual SUBFIND 
(gravitationally bound) subhalos have formed \citep{Vogelsbergeretal2014a, Vogelsbergeretal2014b}. The present time, at $z = 0$, corresponds to snapshot or 
snap 135.\footnote{136 data snapshots are stored for each Illustris run, the first one, snap 0, is at $z = 46.773$.}
We use the subhalo catalog to get the dark matter mass, $M_{\rm dm}$, and the 
stellar mass, $M_{\rm star}$, of each SUBFIND subhalo and find that most 
of them have zero stellar mass.\footnote{These subhalo masses are 
Illustris SubhaloMassType masses. These are the total masses of 
all member particle/cells which are bound to the subhalo under consideration, 
separated by type. SubhaloMassType does not account for particle/cells bound to 
substructures of the subhalo under consideration. Later in this paper we will
also use two other mass definitions.} 
Ignoring these leaves 307,786 subhalos, about 
7.0\% of all subhalos. We then select subhalos with $M_{\rm dm}<M_{\rm star}$, 
finding 4216 subhalos which have more stellar mass than dark matter mass, 
about 1.3\% of the subhalos that have nonzero stellar mass. Most of these 
have low stellar mass, less than $10^{10}\,M_\odot$. There are however three 
subhalos with stellar mass greater than $10^{10}\,M_\odot$ with 
$M_{\rm dm}/M_{\rm star} < 0.1$.

\begin{table}
	\centering
	\begin{tabular}{c|c|c|c|c|c|c|c}
		\hline 
		SUBFIND & $M_{\rm tot}$ & $M_{\rm dm}$ & $M_{\rm star}$ & $M_{\rm gas}$ & SFR & $r_{1/2}$ & $r_{\rm ph}$ \\
		ID & ($10^{10}\,M_\odot$) & ($10^{10}\,M_\odot$) & ($10^{10}\,M_\odot$) & ($10^{10}\,M_\odot$) & ($M_\odot/{\rm yr}$) & (kpc) & (kpc) \\
		\hline 
		41124 & 4.45 & 0.260 & 4.18 & 0 & 0 & 2.03 & 5.24 \\ 
		\hline 
		231881 & 1.86 & 0.0726 & 1.79 & 0 & 0 & 1.57 & 2.49 \\
		\hline 
		476171 & 10.54 & 0.915 & 9.42 & 0 & 0 & 1.82 & 4.76\\ 
		\hline 
		41089 & 2645.03 & 2587.20 & 52.94 & 4.67 & 0.0036 & 28.68 & 19.27 \\
		\hline
		231879 & 1752.37 & 1692.51 & 57.81 & 1.85 & 0 & 17.47& 21.83\\
		\hline
		928035 & 0.0659 & 0 & 0.0659 & 0 & 0 & 1.75 & 1.32 \\ \hline
	\end{tabular} 
	\caption{First three data rows list the 
properties of the three massive dark matter deficient subhalos in snap 135 
at $z = 0$. Fourth and fifth data rows give properties of the subhalos that 
subhalos 41124 and 231881 are merging with. Last data row lists properties of
subhalo 928035 which is discussed below. $M_{\rm tot}$, $M_{\rm dm}$, $M_{\rm star}$ and $M_{\rm gas}$ are total mass, dark matter mass, stellar mass, and gas mass of the subhalos. SFR is the star formation rate, $r_{1/2}$ is the half-stellar-mass radius (the radius containing half of the stellar mass of this subhalo), and $r_{\rm ph}$ is the stellar photometric radius (the radius at which the surface brightness profile, computed from all member stellar particles, drops below the limit of 20.7 mag arcsec$^{-2}$ in the K band).}\label{tab:properties}
\end{table}

\begin{figure}
	\centering
	\includegraphics[width=0.3\linewidth]{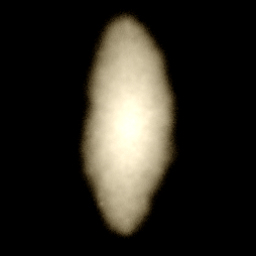}
	\includegraphics[width=0.3\linewidth]{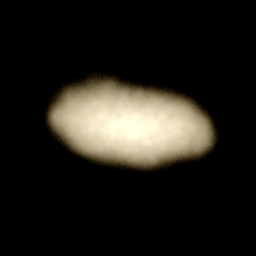}
	\includegraphics[width=0.3\linewidth]{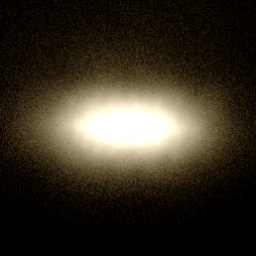}
	\caption{The mock images of the three massive dark matter deficient subhalos. From left to right, these are for 41124, 231881, and 476171 respectively.}
	\label{fig:images}
\end{figure}

\begin{figure}
	\centering
	\includegraphics[width=0.45\linewidth]{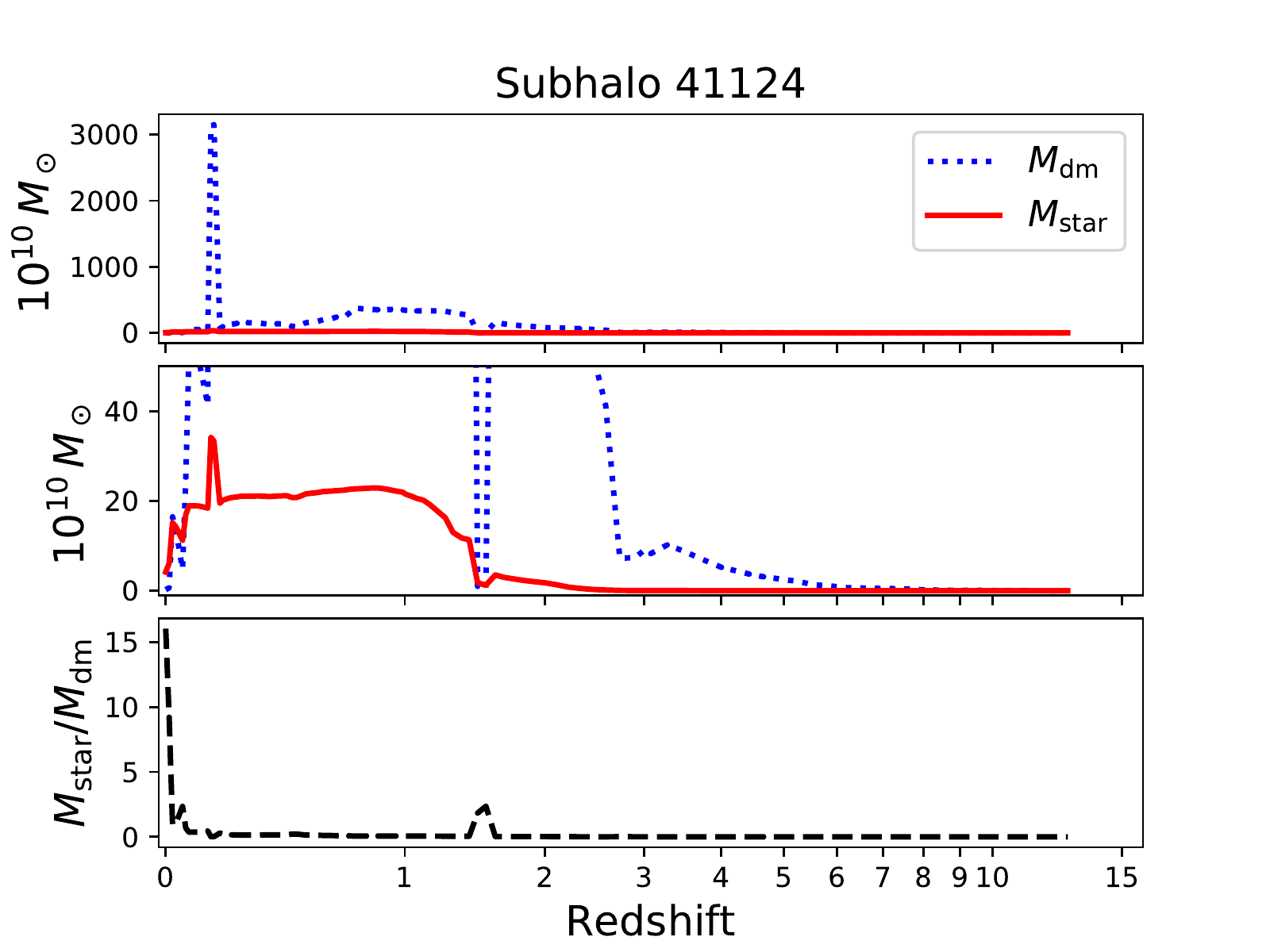}
	\includegraphics[width=0.45\linewidth]{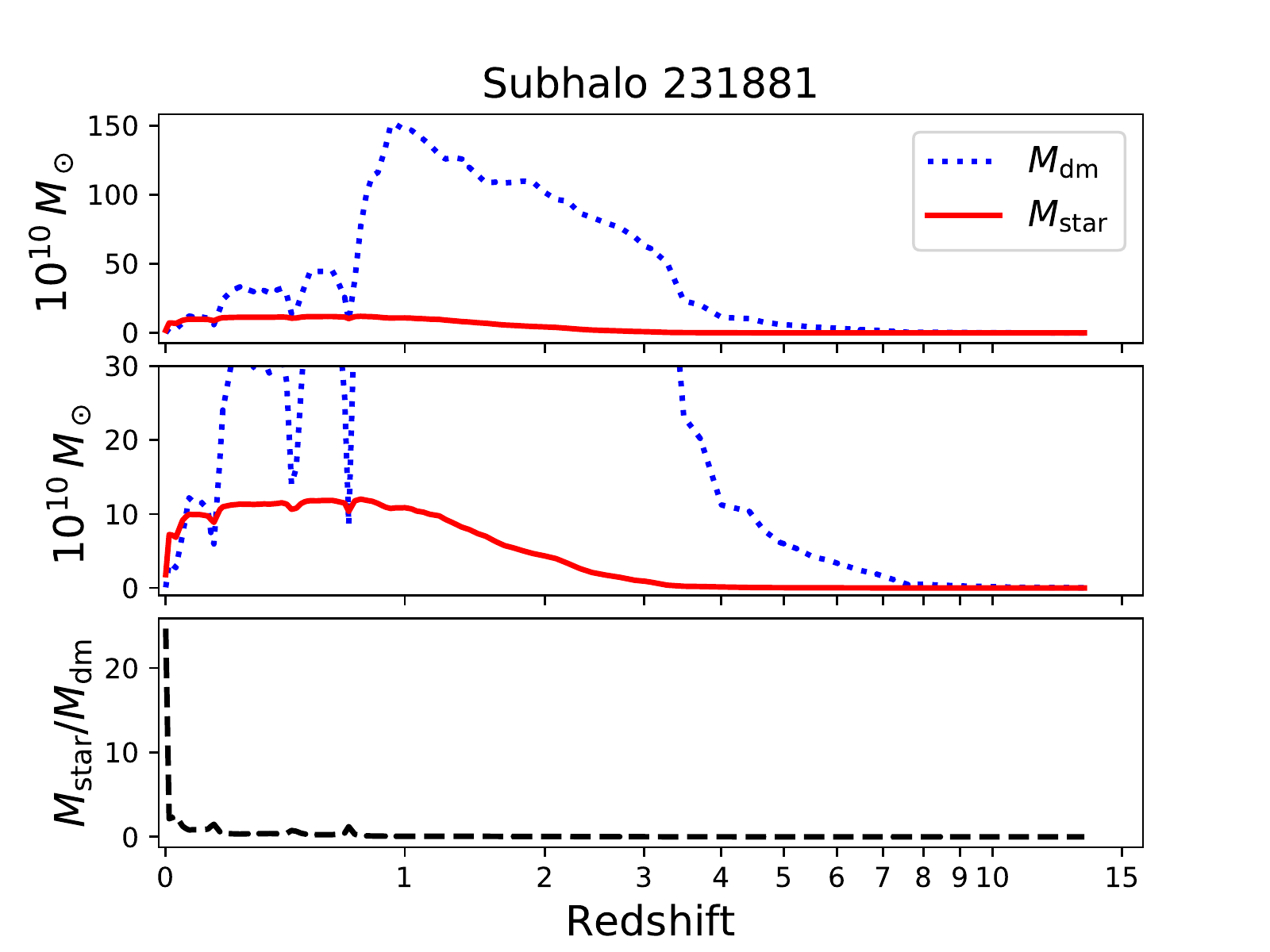}\\
	\includegraphics[width=0.45\linewidth]{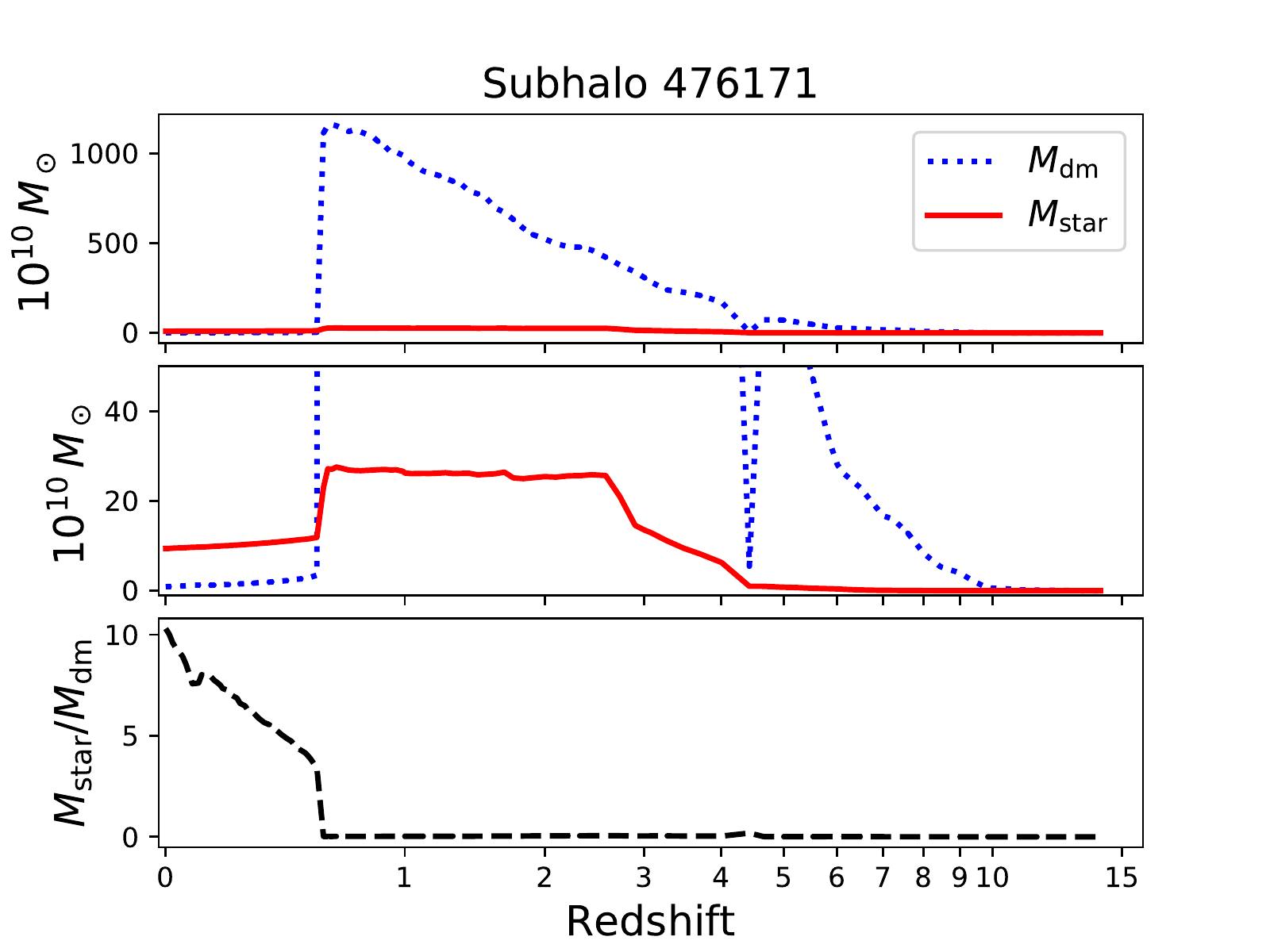}
	\caption{The mass history of the three massive dark matter deficient subhalos. Top subpanels are for dark matter and stars, middle subpanels zoom in on the lower mass region of the top subpanels, and bottom subpanels show the ratio $M_{\rm star}/M_{\rm dm}$. Note the drop in $M_{\rm star}$ and the deeper drop in $M_{\rm dm}$ at low redshift in the upper two subpanels for each of the three subhalos. This results in the relatively rapid increase in $M_{\rm star}/M_{\rm dm}$ for the subhalos at low redshift, seen in the lowest subpanel for each of the three subhalos.}
	\label{fig:mhs}
\end{figure}

The SUBFIND IDs of these three massive dark matter deficient subhalos are 
41124, 231881 and 476171 in snap 135 at $z = 0$ and their properties are 
listed in Table \ref{tab:properties}. We see that these subhalos have no gas 
and their star formation has ceased. Consistent with these properties, the mock 
images of these three subhalos show that they are elliptical subhalos (see 
Fig.\ \ref{fig:images}). We trace the evolutionary history of these subhalos 
by using their sublink merger trees. We plot their dark matter mass and 
stellar mass histories in Fig.\ \ref{fig:mhs}.\footnote{The histories plotted 
in Fig.\ \ref{fig:mhs} are assembled from that of the identified subhalo as
well as its progenitor subhalo(s).}
All three subhalos have a period during which they rapidly lose most of their 
dark matter mass but lose relatively little of their stellar mass which 
results in abrupt increases of their $M_{\rm star}/M_{\rm dm}$ ratios. This might 
happen in a merger process of two large subhalos or through some other 
mechanism. After checking the evolutionary history of the 
three subhalos we find that both 41124 and 231881 are undergoing mergers. 
Subhalo 41124 is merging with subhalo 41089 and subhalo 231881 is merging with 
subhalo 231879 and the properties of these two new subhalos are also listed in 
Table \ref{tab:properties}. Subhalos 41089 and 231879 are much more massive 
than 41124 and 231881 and so dominate the merger processes and the two dark 
matter halos. If subhalos 41124 and 231881 can escape from their much more 
massive companions, they might form dark matter deficient galaxies. Subhalo 
476171, whose stellar mass is 9.42$\times10^{10}\,M_\odot$, comparable to that 
of our Milky Way, is an isolated elliptical galaxy and we will more closely 
examine its evolutionary history to understand how it formed.

Not all of the selected 307,786 SUBFIND subhalos with nonzero stellar mass 
are galaxies. Some of them, especially the less massive subhalos, are more 
likely to be substructures of other galaxies. To find subhalos that are 
galaxies, we check the `Parent' property of each subhalo (the index in 
the subhalo table of the unique SUBFIND host subhalo of this 
subhalo\footnote{This index is local to the FoF group. For example, index 2 
indicates the third most massive subhalo of the parent halo of this subhalo, 
not the third most massive of the whole snapshot.}) which 
can indicate if the subhalo of interest belongs to (is substructure in) a 
host subhalo. However, even if the value of the Parent index is 0 this could 
just mean that the host of the subhalo under consideration is just the 
most massive subhalo in the FoF group of the subhalo under consideration and
not a separate galaxy. So we additionally compare the distance $d$ between 
the subhalo under consideration and its host subhalo with the 
half-stellar-mass radius $r_{1/2}$, which contains half of the total 
stellar mass of the galaxy subhalo, of the host subhalo and regard the subhalo
under consideration to be a galaxy only if it has $d>2r_{1/2}$. We also discard
subhalos with fewer than 20 stellar particles since half-stellar-mass radius
is not measured as robustly for such objects. Applying these
criteria, we get a new galaxy subhalo sample 
containing 58,025 subhalos that are galaxies. We will analyze this 
compilation in what follows.

\begin{figure}
	\centering
	\includegraphics[width=0.8\linewidth]{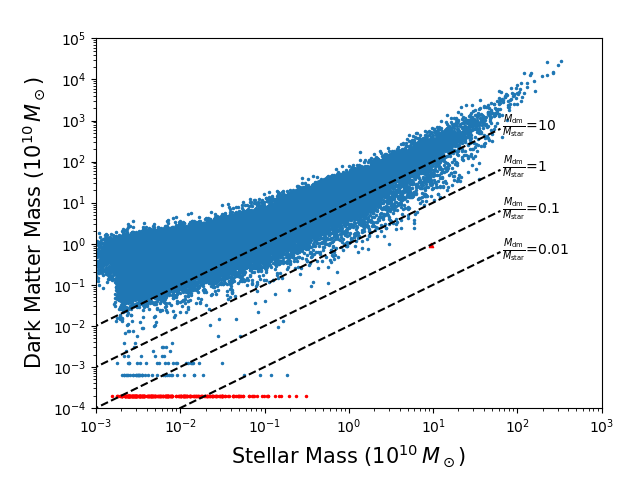}
	\caption{The distribution of the galaxy subhalo sample, with $M_{\rm star}>10^7\,M_\odot$, in the stellar mass---dark matter mass plain. Blue points indicate subhalos and the inclined black dashed lines are loci of the ratio of subhalo dark matter to stellar mass. The lowest horizontal line of blue points correspond to subhalos with one dark matter particle and the lowest horizontal line of red points correspond to those subhalos without dark matter, moved here 
from $M_{\rm dm} = 0$. The red star represents the massive subhalo 476171  ($M_{\rm star} \approx 10^{11}\,M_\odot$) with $M_{\rm dm}/M_{\rm star}<0.1$. (Masses here are SubhaloMassType masses.)}
	\label{fig:massdistribution}
\end{figure} 

Figure \ref{fig:massdistribution} shows the distribution of these galaxy 
subhalos as a function of their stellar mass and dark matter mass. Here we 
consider only the 57,945 galaxy subhalos with stellar mass larger than 
$10^7\,M_\odot$. From the figure we find that there are some subhalos with 
$M_{\rm dm}/M_{\rm star}$ less than 1 which means these subhalos have more 
stellar mass than dark matter mass. There are 420 $M_{\rm dm}/M_{\rm star} < 1$ 
subhalos, which leads to a probability of about 0.72\% for these 
$M_{\rm star} > 10^7\,M_\odot$ dark matter deficient subhalos. It is also clear 
that most of these have low stellar mass, less than 
$10^{10}\,M_\odot$.\footnote{It is unclear what the galaxy subhalos with no 
dark matter (under SubhaloMassType), those indicated by the low red line of 
red points in Fig.\ 
\ref{fig:massdistribution}, are. A few of the more massive ones have 
$M_{\rm star} > 10^9\,M_\odot$. We have looked at a few of the most massive of 
these but they only appear at $z = 0$ (snap 135) and so we are unable to 
trace their evolutionary history. They also do not have image data. It would 
be useful to understand how these galaxy subhalos formed and what 
observational constraints can be placed on them.}  
However, there is one subhalo with stellar mass greater than 
$10^{10}\,M_\odot$ and with $M_{\rm dm}/M_{\rm star} < 0.1$, shown as the red star 
in Fig.\ \ref{fig:massdistribution}. This is subhalo 476171, discussed above,  
which we consider in more detail below. We also find that the other two 
dark matter deficient massive subhalos 41124 and 231881 discussed above 
do not appear in this figure since they are participating in mergers and so 
are regarded as substructures of their much more massive companion galaxies.
We emphasize that galaxy subhalo 476171 is an outlier, and also that it has 
many less extreme --- but still very interesting --- cousins with  
$M_{\rm star} > 10^{10}\,M_\odot$ and $M_{\rm dm}/M_{\rm star} < 1$.

\begin{figure}
\centering
\includegraphics[width=0.8\linewidth]{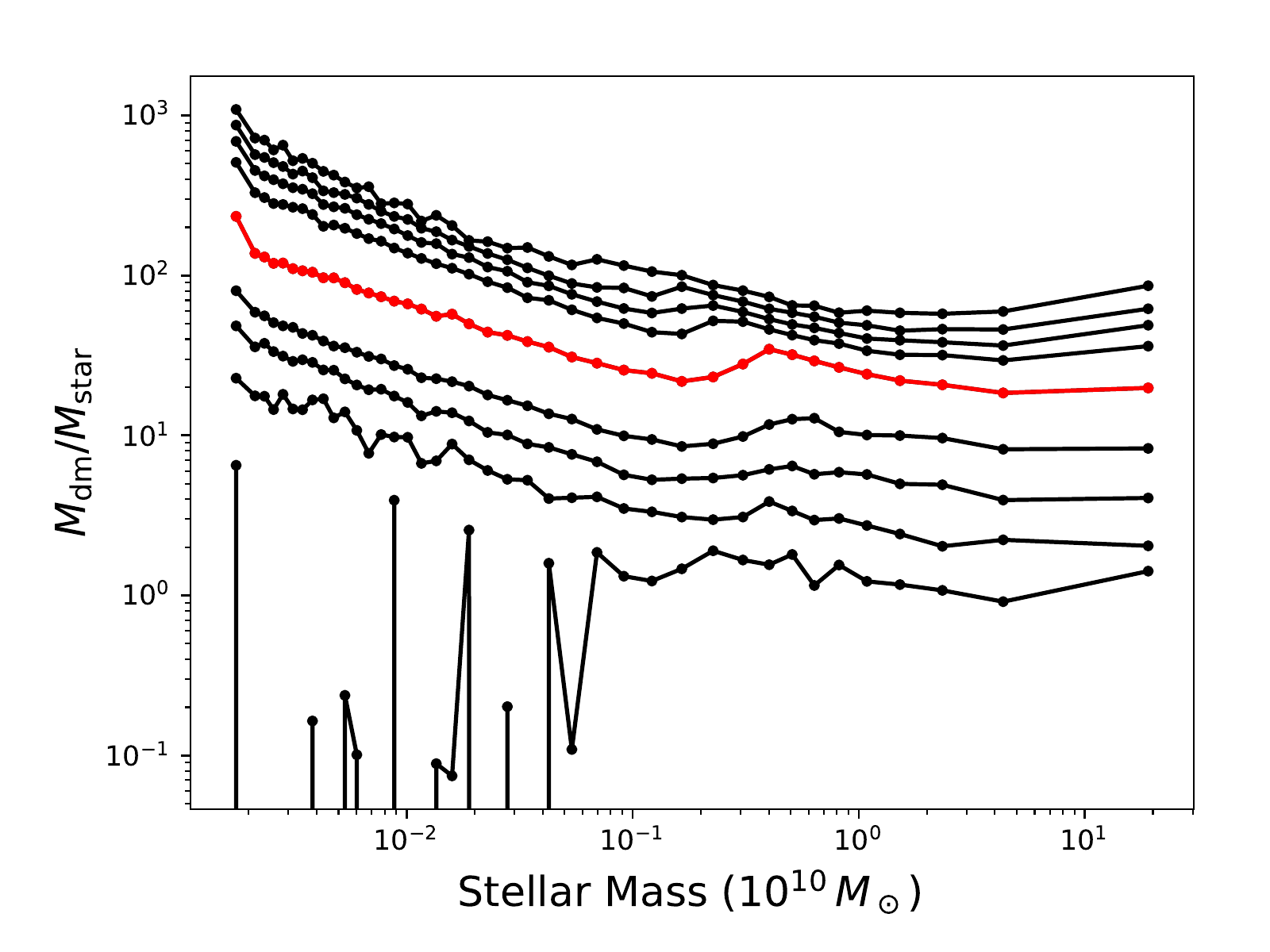}
\caption{The ratio of dark matter mass to stellar mass as a function of stellar mass for the selected galaxy subhalo sample. (Masses here are SubhaloMassType masses.) The central red curve with dots is the median value and the outer black curves with dots represent the upper and lower 1$\sigma$, 1.5$\sigma$, 2$\sigma$, and 2.5$\sigma$ confidence limits. The dots represent the mean stellar mass values in the bins. There are 40 bins and about 1450 objects in each bin.}
	\label{fig:relationofdmsm}
\end{figure}

The relation between the $M_{\rm dm}/M_{\rm star}$ and  $M_{\rm star}$ values of 
the galactic subhalos is shown in Fig.\ \ref{fig:relationofdmsm}. From this 
figure we see that the median value of $M_{\rm dm}/M_{\rm star}$ spans a wide 
range, from about 20 to almost 300, and that this ratio decreases with 
increasing subhalo stellar mass. For given galaxy subhalo stellar mass, this 
ratio also spans a wide range, more so at the low stellar mass end. We see 
that the $M_{\rm dm}/M_{\rm star}$ lower 2.5$\sigma$ limit curve lies at 
$M_{\rm dm}/M_{\rm star}\approx$ 1---3 at the higher stellar mass end and, 
excluding some isolated bins, drops precipitously below about 
$M_{\rm star}=6\times10^8\,M_\odot$. This result means that in the 
$\Lambda$CDM model there is some probability of forming galactic subhalos 
without or with only very little dark matter mass.

\begin{figure}
	\centering
	\includegraphics[width=0.7\linewidth]{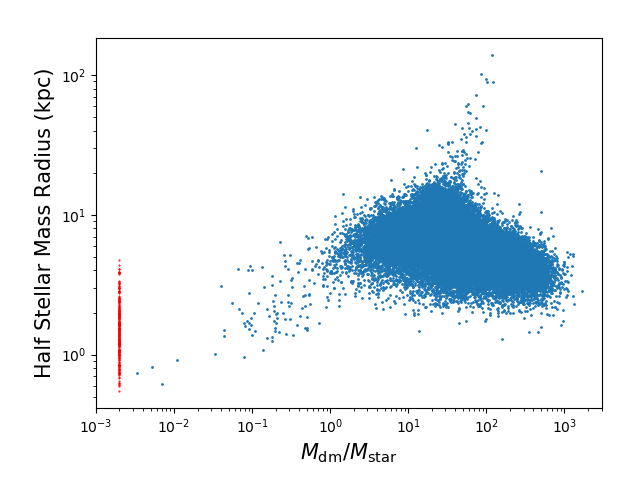}
	\caption{The distribution of the 57,945 galaxy subhalos with $M_{\rm star} > 10^7\,M_\odot$ in the $r_{1/2}$---$(M_{\rm dm}/M_{\rm star})$ plain. (Masses here are SubhaloMassType masses.) The left red line of red dots represent those subhalos with $M_{\rm dm}/M_{\rm star}=0$, moved here from $M_{\rm dm}/M_{\rm star} = 0$. There are a number of galaxy subhalos in the $r_{1/2}>$ 2 kpc and $M_{\rm dm}/M_{\rm star}<1$ region; perhaps these might be called ultra-diffuse galaxies. We note the interesting galaxy subhalos with $r_{1/2}>$ 50 kpc and $M_{\rm dm}/M_{\rm star}> 50$.}
	\label{fig:Rms_r21_sh}
\end{figure}

Since NGC1052-DF2 is said to be an ultra-diffuse galaxy, we also consider 
the half-stellar-mass radius, $r_{1/2}$, and plot the distribution of the 
galaxy subhalos in the $r_{1/2}$---$(M_{\rm dm}/M_{\rm star})$ plain, see 
Fig.\ \ref{fig:Rms_r21_sh}. From this figure we see that, for say 
$M_{\rm dm}/M_{\rm star} \leq 5$, the half-stellar-mass 
radius $r_{1/2}$ decreases with decreasing galaxy subhalo mass ratio 
$M_{\rm dm}/M_{\rm star}$, if we ignore the red dots. Further, for 
$M_{\rm dm}/M_{\rm star}<1$, there are some galaxy subhalos with relatively 
large half-stellar-mass radius, say $>$ 2\,kpc, which might 
appropriately be called ultra-diffuse galaxies. Figure \ref{fig:Rms_r21_sh}
also shows that the Illustris simulation results in a few galaxy
subhalos with $r_{1/2}>$ 50 kpc. We examined their properties and found 
that they are the most or the second most massive galaxy subhalos in massive 
FoF groups, perhaps cD galaxies in galaxy clusters.\footnote{Additionally, 
the left most red line of red dots in Fig.\ \ref{fig:Rms_r21_sh} corresponds 
to galaxy subhalos with 
no dark matter (under SubhaloMassType). Some of these have fairly large 
half-stellar-mass radius $r_{1/2}$, greater than a few kpc. It would 
be useful to understand how these galaxy subhalos formed and what 
observational constraints can be placed on them.} 
It is of interest to better understand these objects and to also 
observationally search for them in the real universe.

As noted above, Illustris SubhaloMassType masses are the total masses of 
all member particle/cells which are bound to the subhalo under consideration, 
separated by type. It is possible that some of the SubhaloMassType dark matter 
deficient galaxy subhalos we have found have diffuse dark matter halos whose 
masses do not contribute to $M_{\rm dm}$ computed using 
SubhaloMassType.\footnote{We thank V.\ Springel for pointing this out and for 
suggesting that we should also examine the raw Illustris data.}
To examine this possibility we use the raw Illustris data of the 420 
$M_{\rm dm}/M_{\rm star} < 1$ and $M_{\rm star} > 10^7\,M_\odot$ 
dark matter deficient galaxy subhalos (here $M_{\rm dm}$ and $M_{\rm star}$ are 
computed using SubhaloMassType).

\begin{figure}
	\centering
	\includegraphics[width=0.48\linewidth]{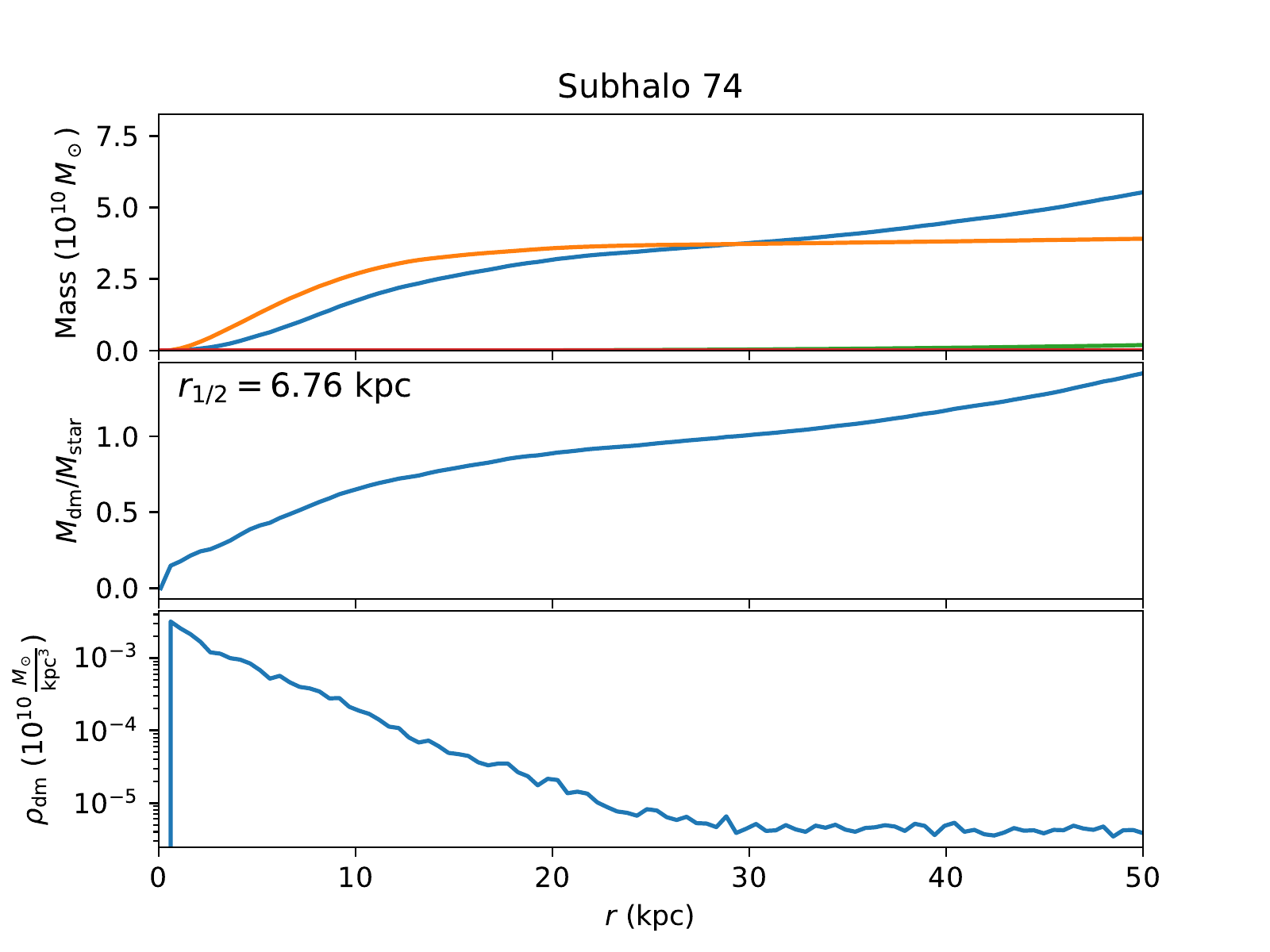}
	\includegraphics[width=0.48\linewidth]{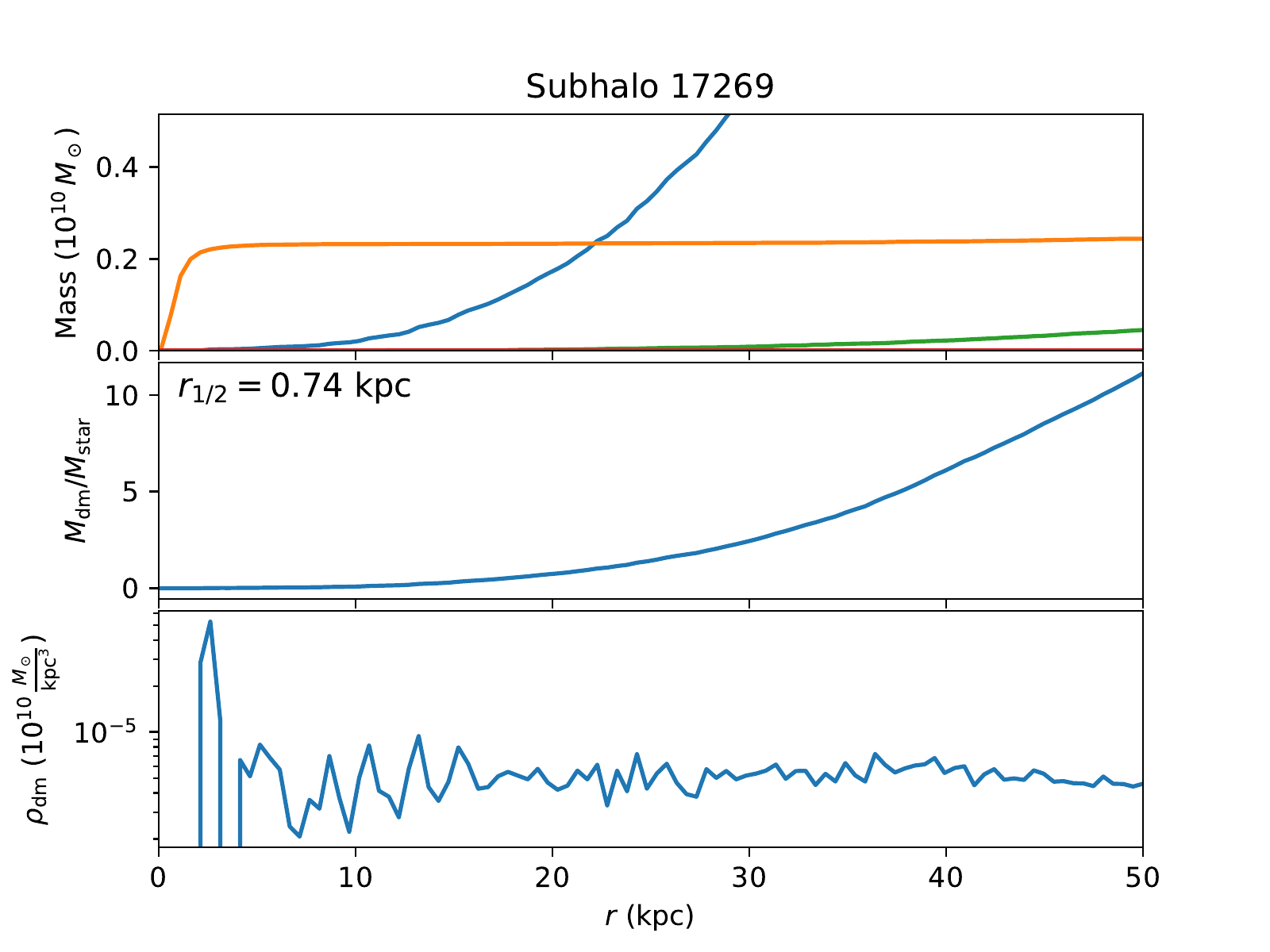}\\
	\includegraphics[width=0.48\linewidth]{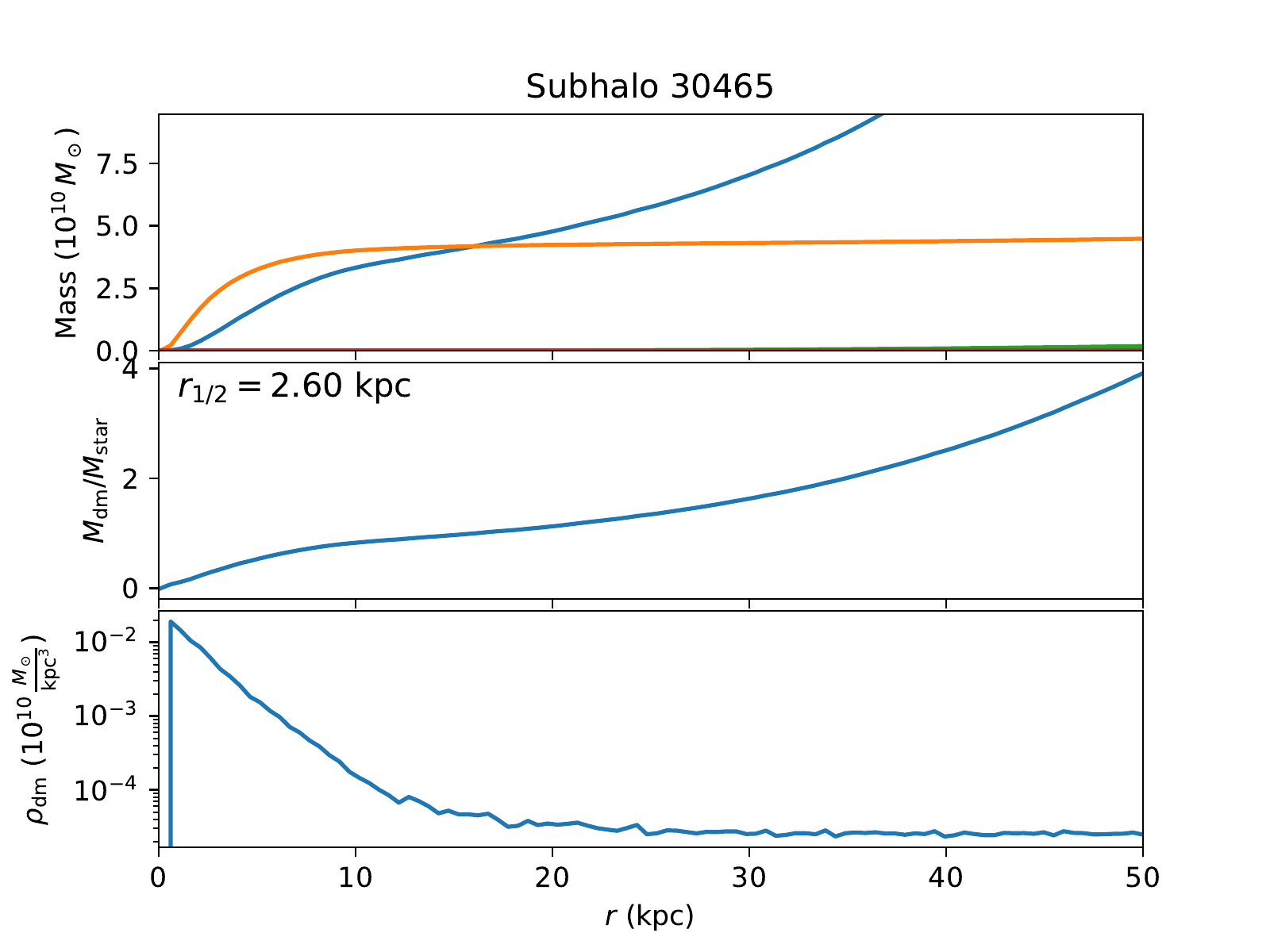}
	\includegraphics[width=0.48\linewidth]{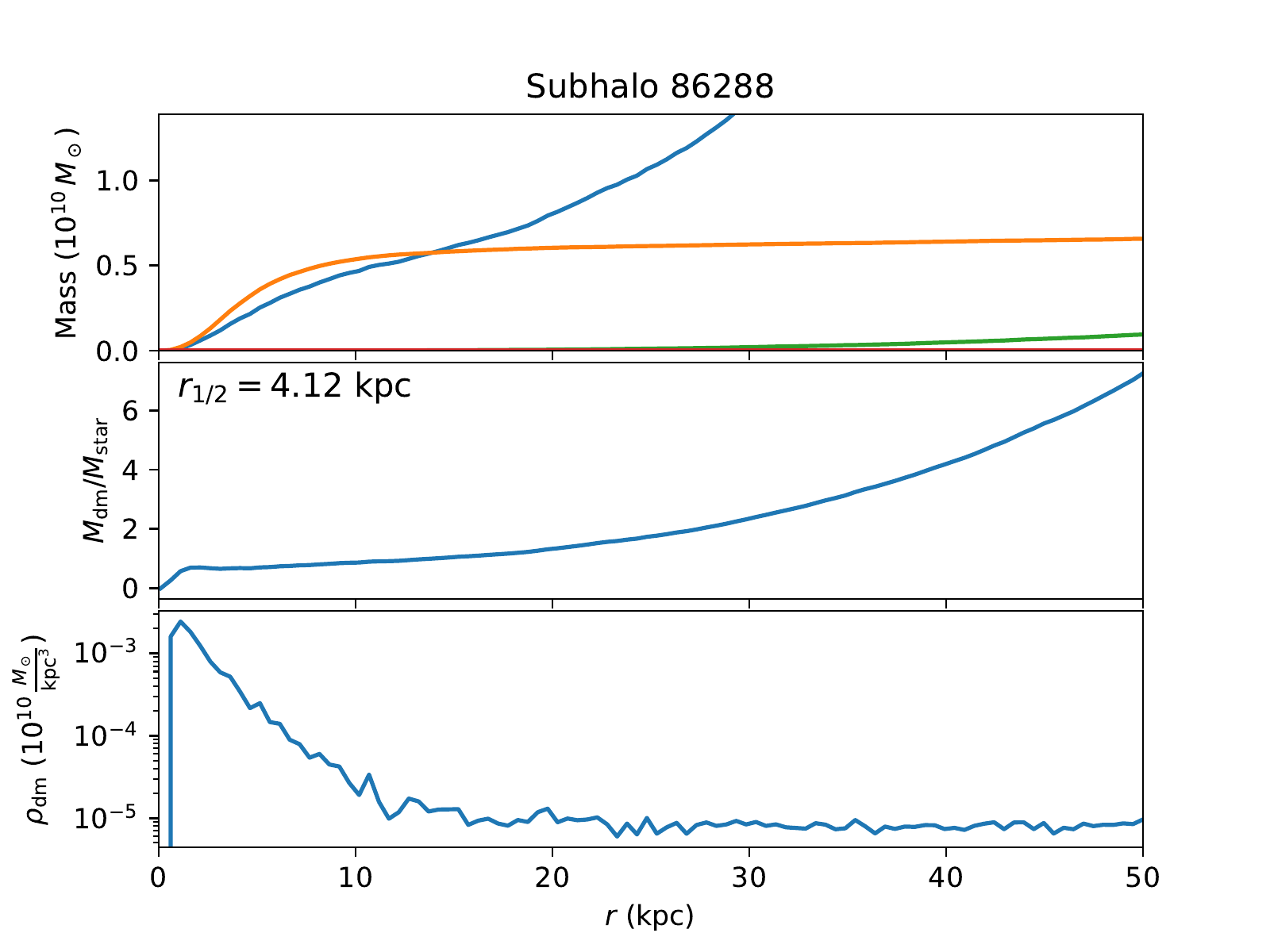}\\
	\includegraphics[width=0.48\linewidth]{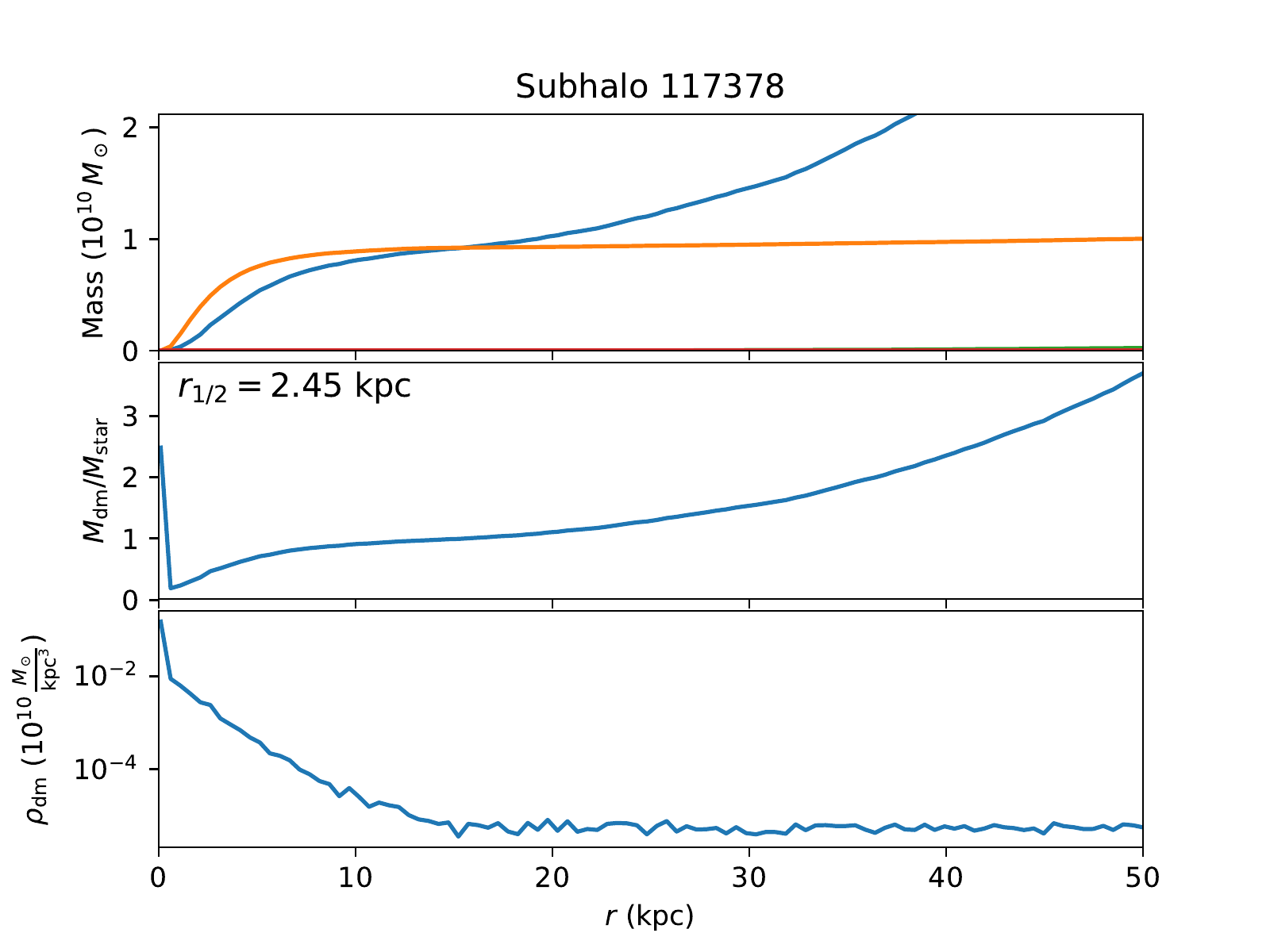}
	\includegraphics[width=0.48\linewidth]{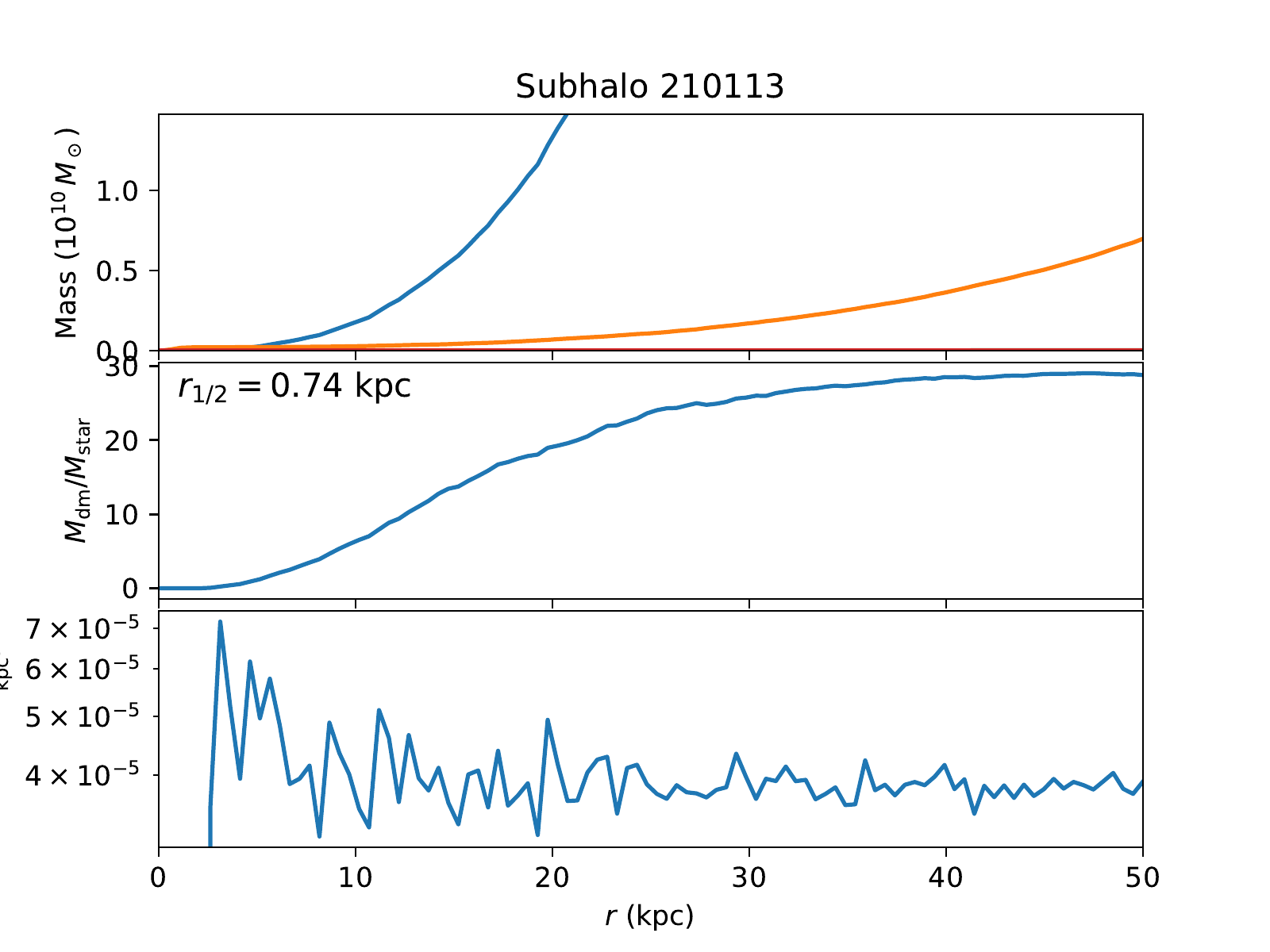}\\
	\caption{Density profiles of different particles for some dark matter deficient galaxy subhalos. Each panel title is the subhalo ID, and in each of the six panels the top subpanel shows the various masses within $<r$ (linestyles are defined in the top subpanel of the top-left panel), the middle subpanel shows the $M_{\rm dm}/M_{\rm star}$ within $<r$, and the bottom subpanel shows the density of dark matter at radius $r$. $r_{1/2}$ values listed in the middle subpanels are half-stellar-mass radii computed using SubhaloMassType.}
	\label{fig:dprofile1}
\end{figure}

\begin{figure}
	\centering
	\includegraphics[width=0.48\linewidth]{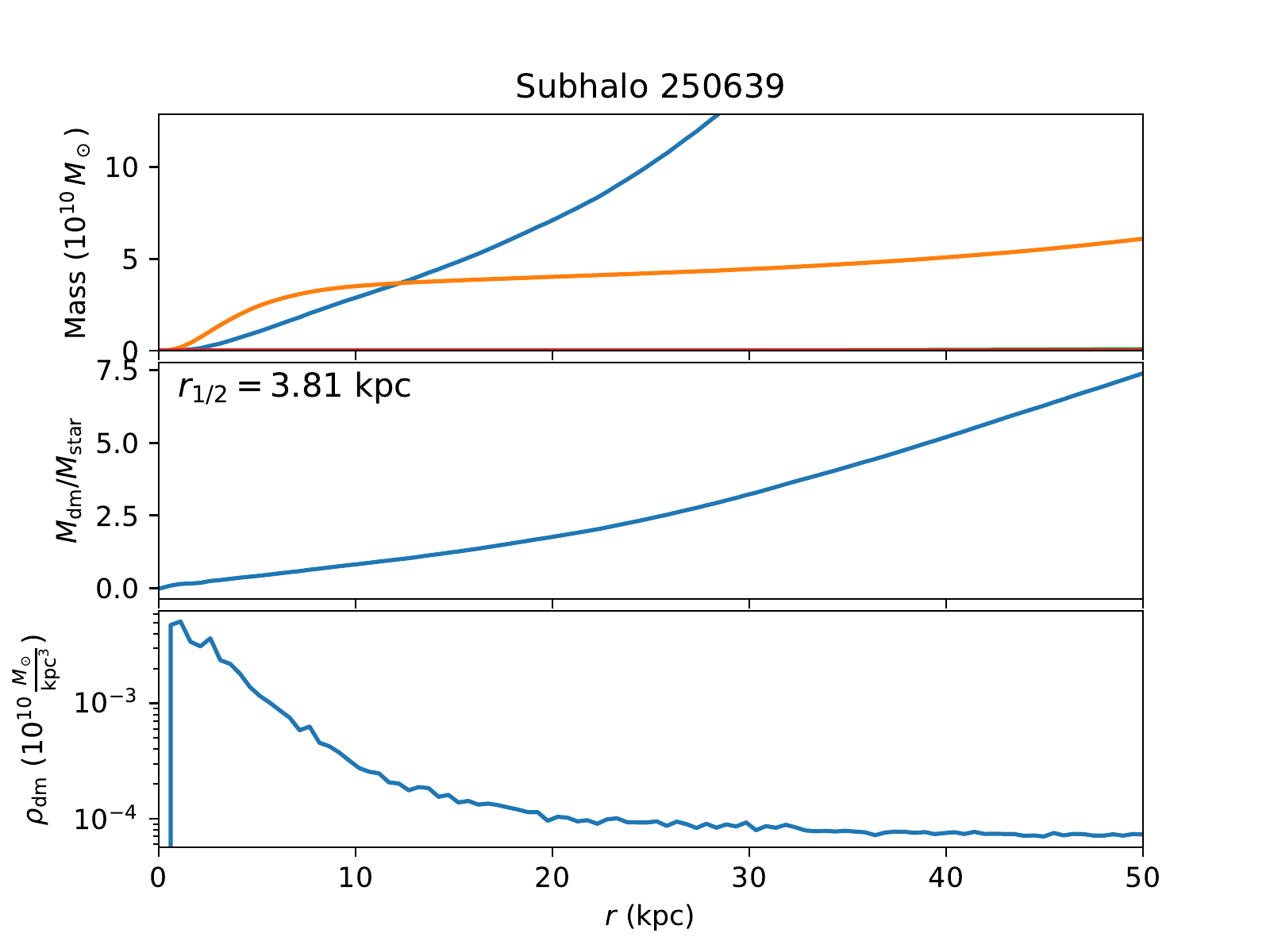}
	\includegraphics[width=0.48\linewidth]{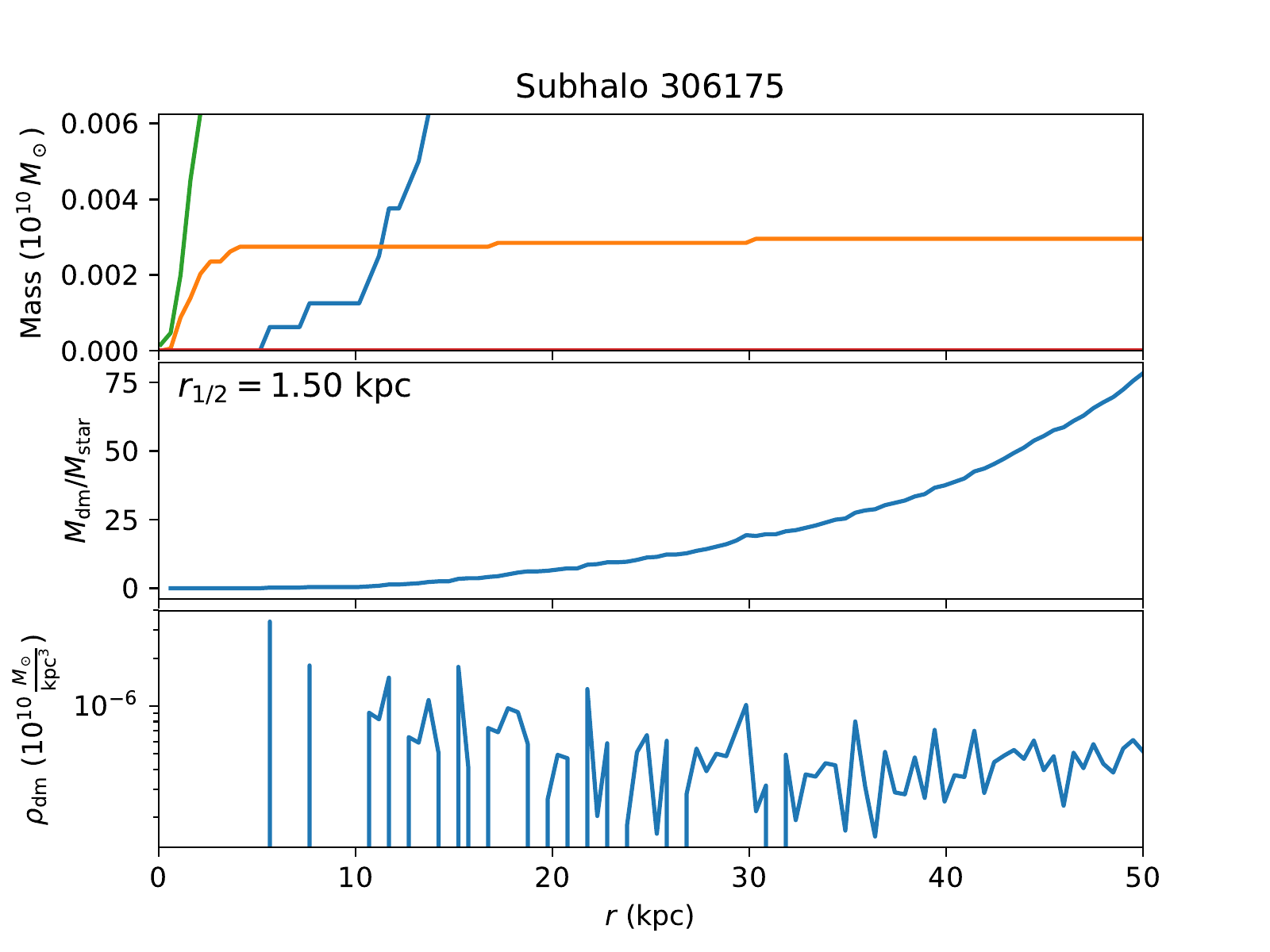}\\
	\includegraphics[width=0.48\linewidth]{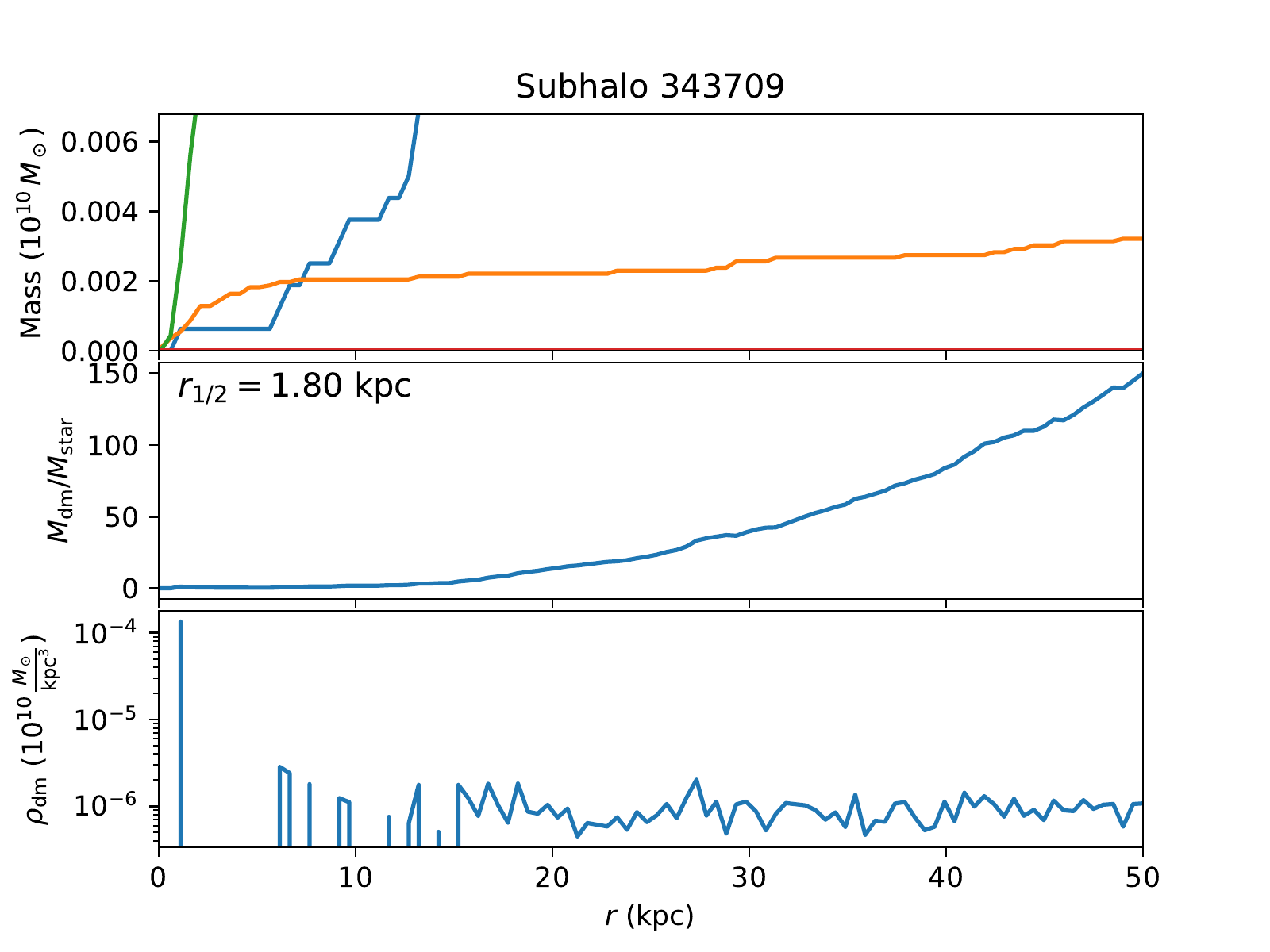}
	\includegraphics[width=0.48\linewidth]{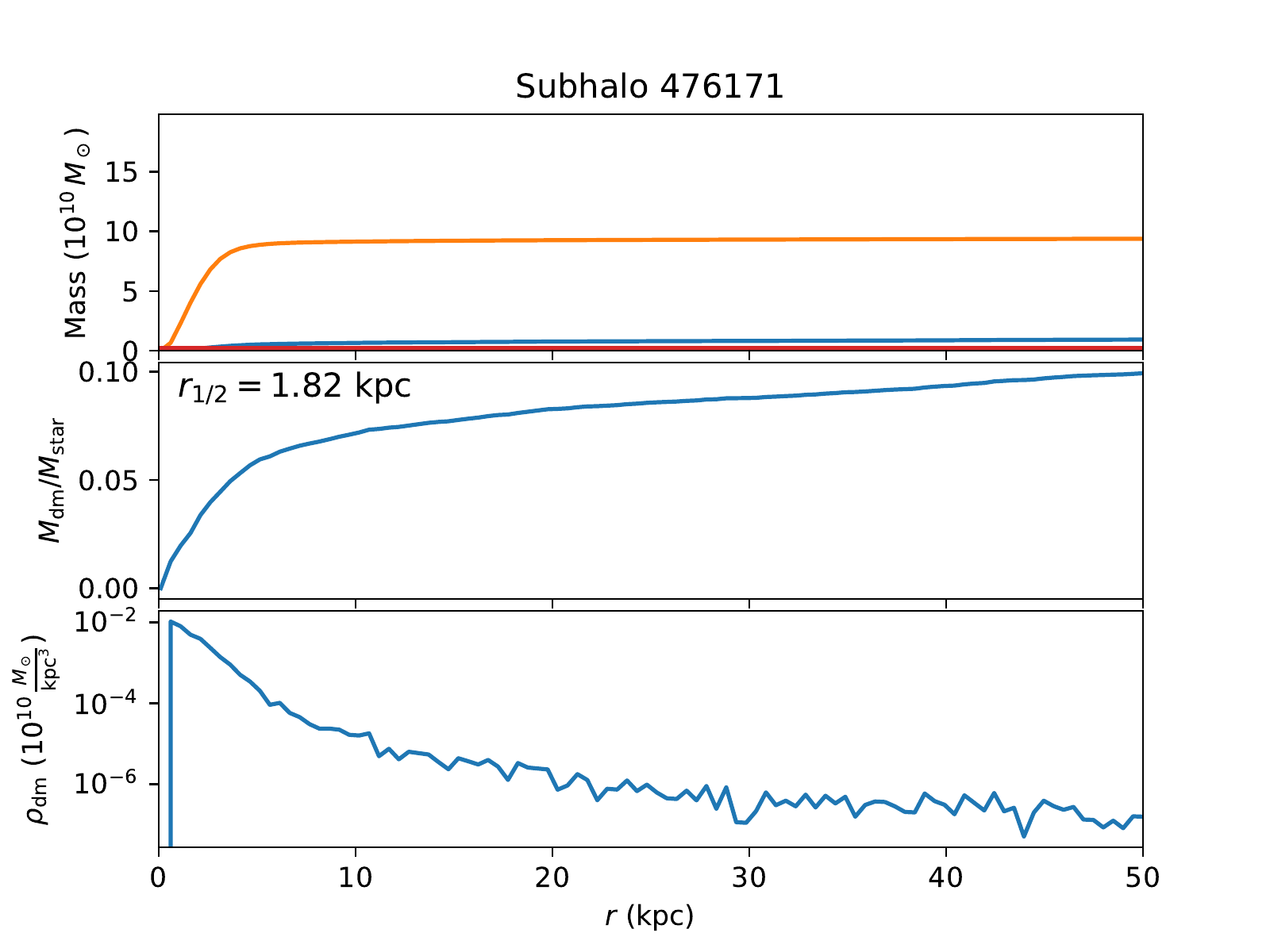}\\
	\includegraphics[width=0.48\linewidth]{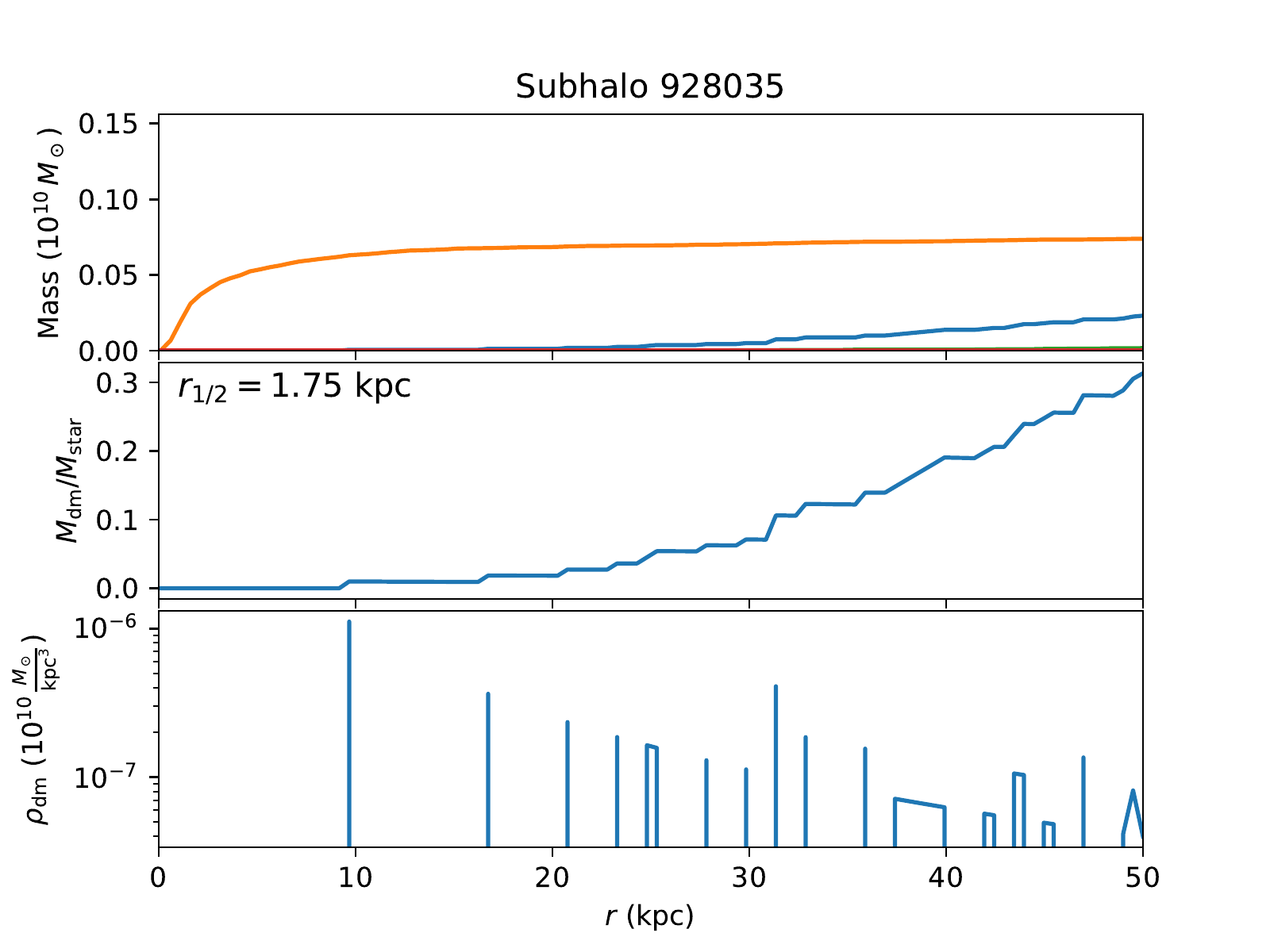}
	\includegraphics[width=0.48\linewidth]{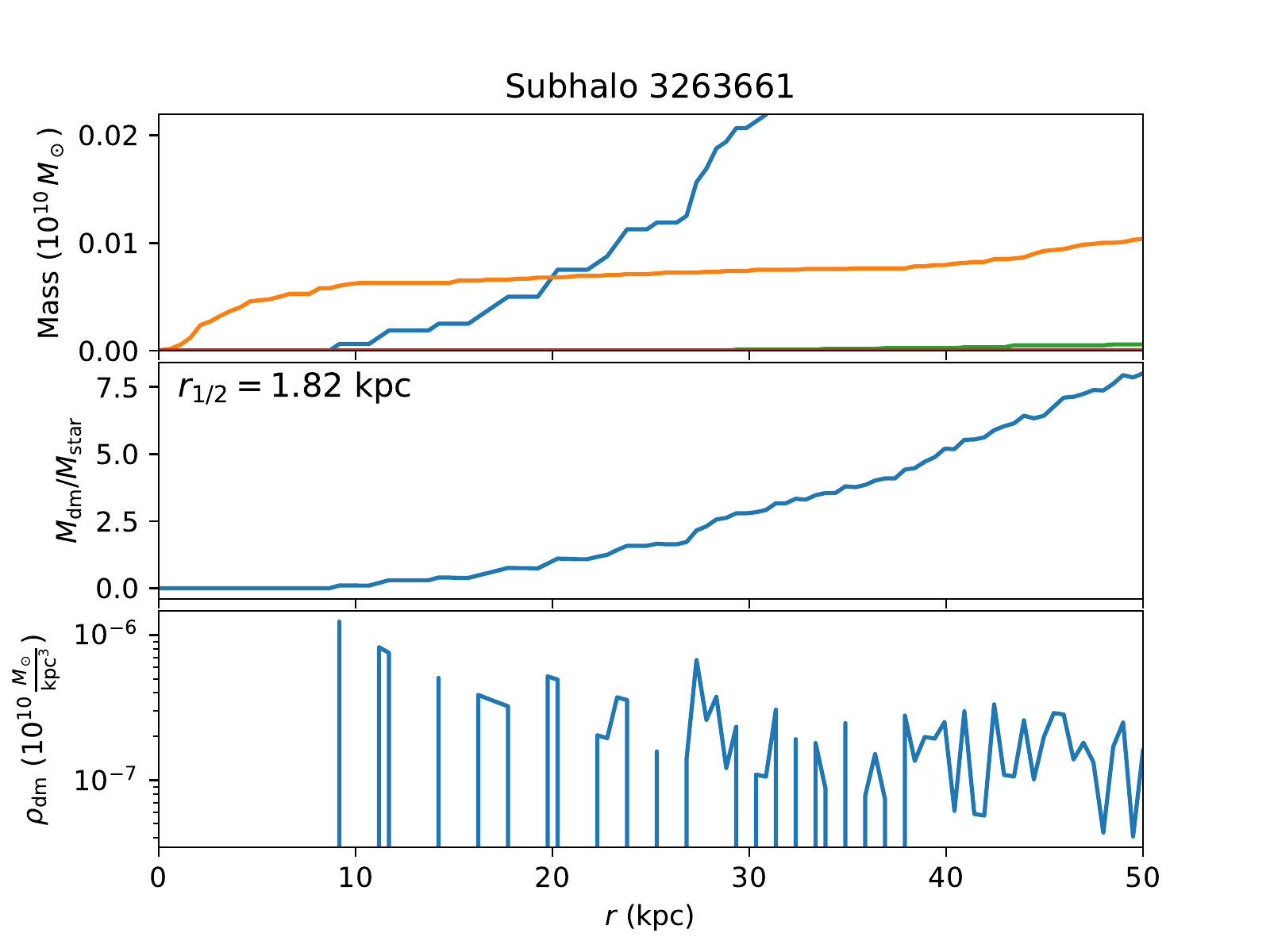}\\
	\caption{See Fig.\ \ref{fig:dprofile1} caption.}
	\label{fig:dprofile2}
\end{figure}

From the raw data for each of the 420 galaxy subhalos we compute spherically 
symmetrized radial density profiles for the dark matter (dm), star, gas, and 
black hole (bh) particle distributions out to a radius of 50 kpc. See Figs.\ 
\ref{fig:dprofile1} and \ref{fig:dprofile2} for some examples. Most of these
420 galaxy subhalos have diffuse dark matter halos.

Only two (of the 420) have $M_{\rm dm}/M_{\rm star}<1$ when the density profiles 
are integrated to a radial distance of 50 kpc from the galaxy subhalo center
to determine $M_{\rm dm}$ and $M_{\rm star}$.
These are subhalos 476171 (center panel in the right hand column of 
Fig.\ \ref{fig:dprofile2}) and 928035 (bottom panel in the left hand column of 
Fig.\ \ref{fig:dprofile2}). We examined subhalo 928035 and found 
that it was an isolated small subhalo first seen at snap 135 ($z = 0$). Its 
main properties are listed in the last data row of Table \ref{tab:properties}. 
Since subhalo 928035 first appears at snap 135, it has no progenitor and 
we are unable to trace its history back in time and so do not know how it 
formed. We discuss the formation history of subhalo 476171 below. 

\begin{figure}
	\centering
	\includegraphics[width=0.7\linewidth]{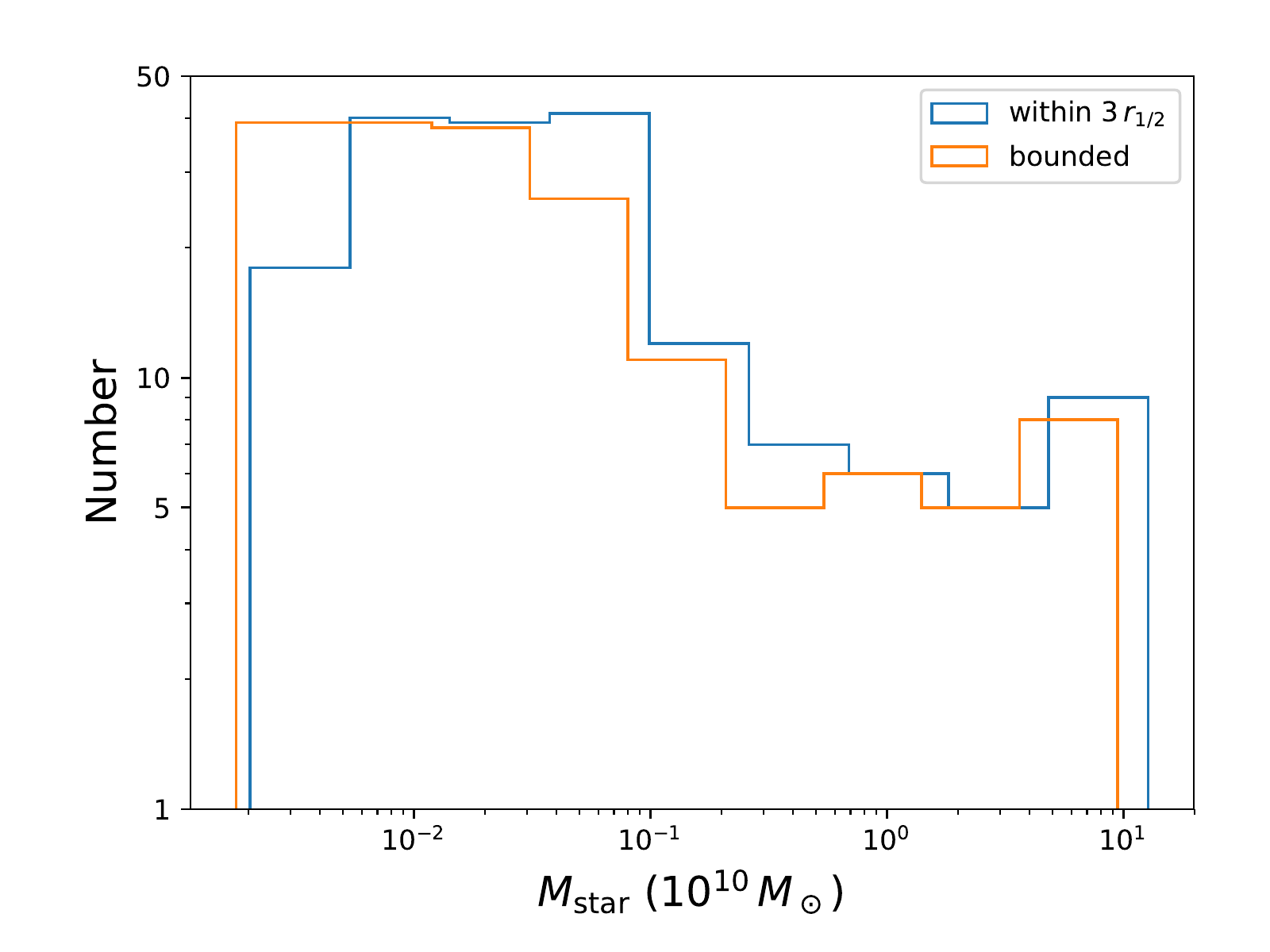}
	\caption{Distribution of the 177 dark matter deficient galaxy subhalos as a function of stellar mass. Blue histogram is for $M_{\rm star}$ computed by integrating the star particle density profile to 3$r_{1/2}$ while the orange histogram is for SubhaloMassType $M_{\rm star}$.}
	\label{fig:177massdistribution}
\end{figure}

\begin{figure}
	\centering
	\includegraphics[width=0.7\linewidth]{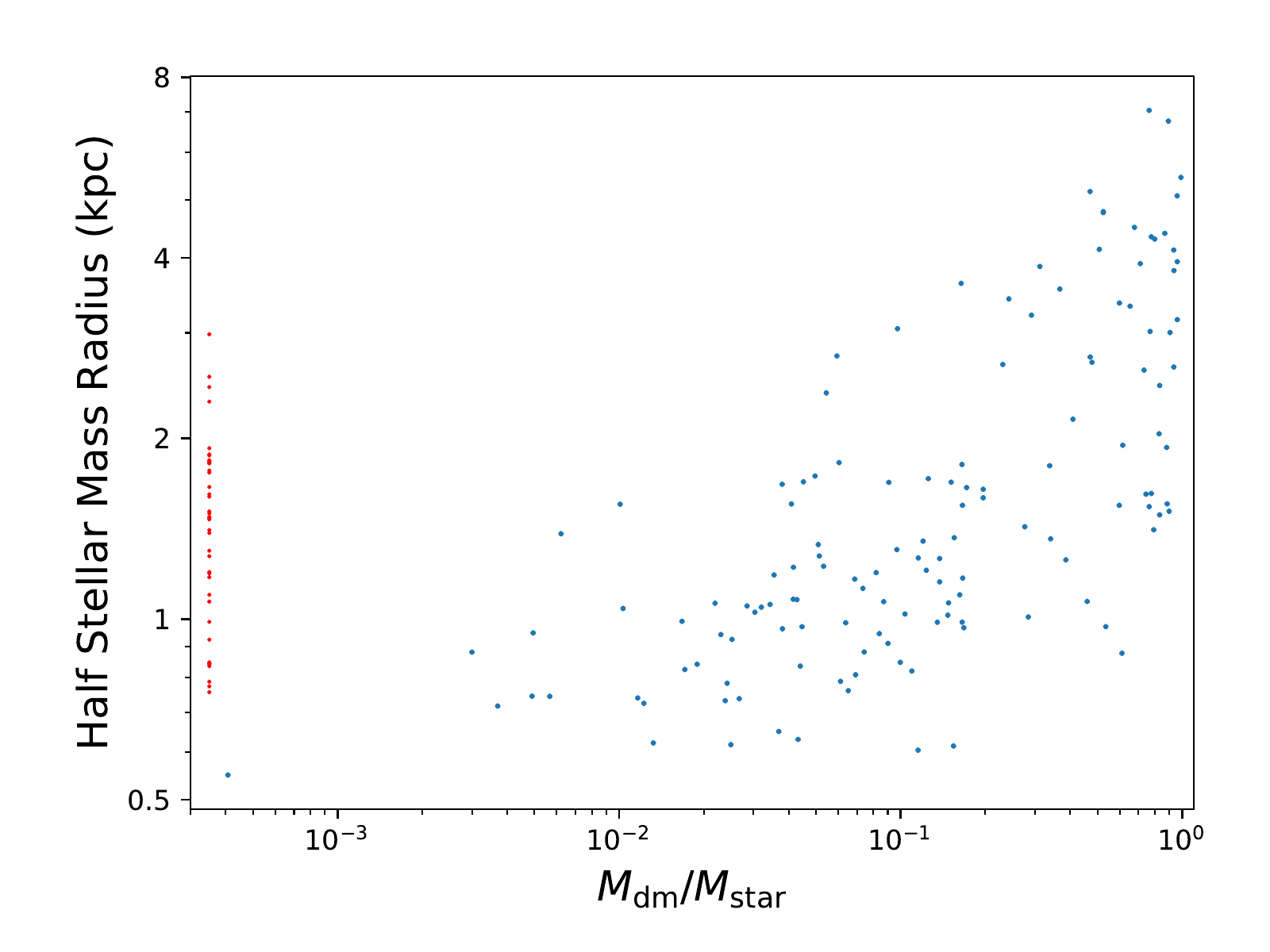}
	\caption{Same as Fig.\ \ref{fig:Rms_r21_sh} but now for only the 177 dark matter deficient galaxy subhalos, with $M_{\rm dm}/M_{\rm star}$ determined from an integration of the density profiles to 3$r_{1/2}$. Perhaps the 47 (26.6\% of 177) galaxy subhalos in the $r_{1/2}>$ 2 kpc and $M_{\rm dm}/M_{\rm star}<1$ region might be called ultra-diffuse galaxies. Note that there still are some galaxy subhalos with no dark matter (the red dots), although many fewer than those in Fig.\ \ref{fig:Rms_r21_sh}.}
	\label{fig:Rms_r21_sh2}
\end{figure}

Observationally, 50 kpc is not necessarily an appropriate distance to 
integrate to to determine whether a galaxy subhalo is dark matter deficient. 
Perhaps more reasonable is to integrate out to a small multiple of the 
galaxy subhalo half-stellar-mass radius $r_{1/2}$.\footnote{We use $r_{1/2}$ 
computed using SubhaloMassType as we are interested only in the star particles
bound to the subhalo.} 
If we integrate to 3$r_{1/2}$ we find that 177 (of the 420 SubhaloMassType 
$M_{\rm dm}/M_{\rm star} < 1$ and $M_{\rm star} > 10^7\,M_\odot$) galaxy subhalos
have raw data density profile $M_{\rm dm}/M_{\rm star} < 1$ and so are dark matter 
deficient. These 177 galaxy subhalos are 0.31\% of the 57,945 Illustris 
galaxy subhalos with SubhaloMassType $M_{\rm star} > 10^7\,M_\odot$. Masses,
$M_{\rm dm}/M_{\rm star}$'s, and density profiles for a dozen of these 177 subhalo
galaxies are shown in Figs.\ \ref{fig:dprofile1} and \ref{fig:dprofile2}.
The distribution of these 177 subhalo galaxies as a function of stellar mass 
is shown in Fig.\ \ref{fig:177massdistribution}. Figure \ref{fig:Rms_r21_sh2} 
shows the distribution of these 177 galaxy subhalos in the $r_{1/2}$---$(M_{\rm dm}/M_{\rm star})$ plain.

Figure \ref{fig:177massdistribution} shows that the distribution in stellar 
mass of the 177 dark matter deficient galaxy subhalos is reasonably flat. It 
is of significant interest to determine the relative probability of these 
177 subhalos as a function of stellar mass. A proper computation of this 
probability requires a determination of $M_{\rm dm}/M_{\rm star}$ from the 
density profiles of all 57,945 galaxy subhalos with 
$M_{\rm star} > 10^7\,M_\odot$. This is beyond the computer resources we have 
available. We have found that 42\% of the SubhaloMassType 
$M_{\rm dm}/M_{\rm star}<1$ galaxy subhalos are indeed dark matter dominated 
when we integrate the density profiles to 3$r_{1/2}$ 
(177 out of 420).\footnote{This ratio will depend on the distance to which the 
density profiles are integrated, also, for the following $M_{\rm star}$-bins
discussion, the ratio will depend on the value of $M_{\rm star}$ under 
consideration.}  
Consequently $M_{\rm dm}/M_{\rm star}$ determined using SubhaloMassType is 
likely to provide a qualitatively reasonable approximation of the more
correct $M_{\rm dm}/M_{\rm star}$ determined from an integration of the density
profiles and so we use the SubhaloMassType $M_{\rm dm}/M_{\rm star}$'s to 
qualitatively estimate the relative probability (an estimate that should be 
correct to within a factor of 2 or 3).

\begin{figure}
	\centering
	\includegraphics[width=0.7\linewidth]{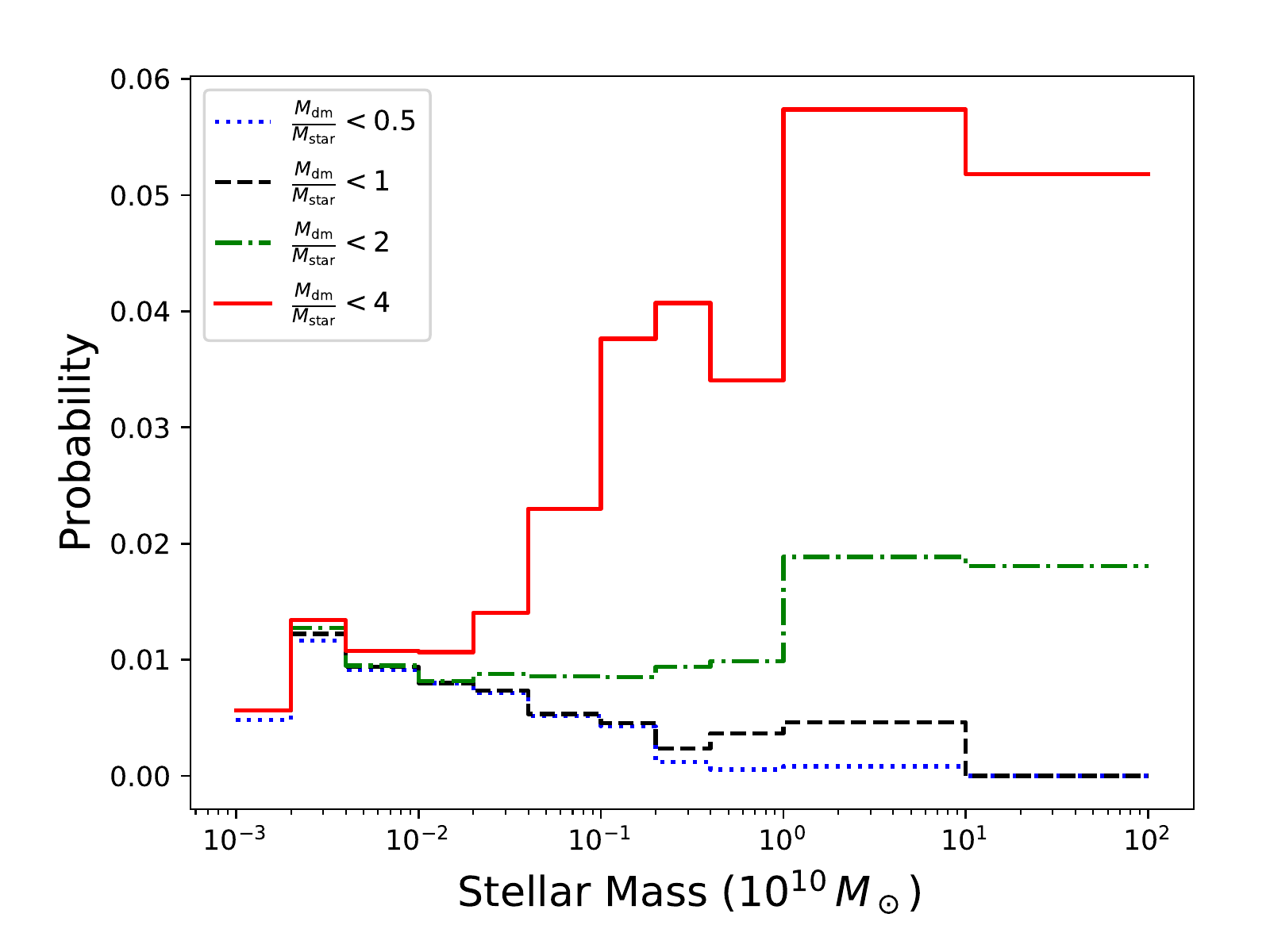}
	\caption{Fractional number of galaxy subhalos satisfying the indicated SubhaloMassType $M_{\rm dm}/M_{\rm star}$ constraint, relative to the total number of subhalos in the corresponding stellar mass bin. See Table \ref{tab:probability} for numerical values.}
	\label{fig:probability}
\end{figure}

Figure \ref{fig:probability} and Table \ref{tab:probability} show and list 
the relative probabilities of dark matter deficient (under SubhaloMassType) 
galaxy subhalos in a 
range of stellar mass bins. We see that more than  0.7\% of galaxy subhalos 
with SubhaloMassType $M_{\rm star} \approx 2 \times 10^8\, M_\sun$ \citep[as found for the ultra-diffuse dwarf galaxy NGC1052-DF2 by][]{vanDokkumetal2018a} have 
SubhaloMassType $M_{\rm dm}/M_{\rm star} < 1$. Even for galaxies with $M_{\rm star} \approx 10^9\, M_\sun$, which should be easier to see, more than 0.4\% have $M_{\rm dm}/M_{\rm star} < 0.5$. If $\Lambda$CDM and the Illustris simulation are accurate, given these SubhaloMassType probabilities, perhaps the more surprising thing about NGC1052-DF2 is not the low $M_{\rm dm}/M_{\rm star}$ but rather that it took so long to find the first dark matter deficient galaxy (if it is indeed confirmed as such). 

\begin{table}
	\centering
	\begin{tabular}{c|c|c|c|c|c|c|c|c|c}
		\hline 
		\multirow{3}*{Stellar mass $(10^{10}\,M_\odot)$ bin} & \multirow{3}*{Total number of subhalos} & \multicolumn{8}{c}{Subhalos with $M_{\rm dm}/M_{\rm star}$} \\
		\cline{3-10}
			&	& \multicolumn{4}{c|}{Number} & \multicolumn{4}{c}{Probability}\\
		\cline{3-10}
			&	& $<0.5$ & $<1$ & $<2$ & $<4$ & $<0.5$ & $<1$ & $<2$ & $<4$ \\ \hline

[1,2)$\times10^{-3}$	&	1240	&	6	&	7	&	7	&	7	&	0.48\%	&	0.56\%	&	0.56\%	&	0.56\%	\\	\hline
[2,4)$\times10^{-3}$	&	10142	&	118	&	124	&	129	&	136	&	1.16\%	&	1.22\%	&	1.27\%	&	1.34\%	\\	\hline
[0.4,1)$\times10^{-2}$	&	10946	&	100	&	103	&	104	&	118	&	0.91\%	&	0.94\%	&	0.95\%	&	1.08\%	\\	\hline
[1,2)$\times10^{-2}$	&	6380	&	51	&	51	&	52	&	68	&	0.80\%	&	0.80\%	&	0.82\%	&	1.07\%	\\	\hline
[2,4)$\times10^{-2}$	&	4909	&	35	&	36	&	43	&	69	&	0.71\%	&	0.73\%	&	0.88\%	&	1.41\%	\\	\hline
[4,10)$\times10^{-2}$	&	5258	&	27	&	28	&	45	&	121	&	0.51\%	&	0.53\%	&	0.86\%	&	2.30\%	\\	\hline
[0.1,0.2)	&	3294	&	14	&	15	&	28	&	124	&	0.43\%	&	0.46\%	&	0.85\%	&	3.76\%	\\	\hline
[0.2,0.4)	&	3414	&	4	&	8	&	32	&	139	&	0.12\%	&	0.23\%	&	0.94\%	&	4.07\%	\\	\hline
[0.4,1)	&	5463	&	3	&	20	&	54	&	186	&	0.05\%	&	0.37\%	&	0.99\%	&	3.40\%	\\	\hline
[1,10)	&	6047	&	5	&	28	&	114	&	347	&	0.08\%	&	0.46\%	&	1.89\%	&	5.74\%	\\	\hline
[10,100]	&	830	&	0	&	0	&	15	&	43	&	0.00\%	&	0.00\%	&	1.81\%	&	5.18\%	\\	\hline

	\end{tabular} 
	\caption{Census of galaxy subhalos in different stellar mass bins. Masses are SubhaloMassType masses.}\label{tab:probability}
\end{table}

We now consider in more detail the massive dark matter deficient galaxy 
subhalo 476171. In snap 135 at $z = 0$, subhalo 476171 is isolated with 
no satellite subhalo or substructure. Its half-stellar-mass radius is 
1.82 kpc and its stellar photometric radius\footnote{The radius at which 
the surface brightness profile, computed from all member stellar particles, 
drops below the limit of 20.7 mag arcsec$^{-2}$ in the K band.} is 
4.76 kpc. Galaxy subhalo 476171 is very small relative to its large stellar 
mass and so the stars in 476171 are bound together
tightly.\footnote{It is possible that at least the more 
massive dark matter deficient galaxy subhalos (with $M_{\rm dm}/M_{\rm star}<1$ 
and $M_{\rm star} > 10^{10}\, M_\sun$, see Fig.\ \ref{fig:massdistribution})
have to be more tightly bound and hence might be expected to be ellipticals. 
Given their small spatial extent they 
should have higher surface brightness than more normal dark matter dominated 
galaxies of the same stellar mass. As a result they will likely not look like 
more normal dark matter dominated galaxies. It is of interest to determine
what observational limits exist on such dark matter deficient galaxies and 
it is also of interest to search for such galaxies.}  
From Table \ref{tab:properties} we see that the other two massive dark matter 
deficient subhalos are also relatively small. 

Galaxy subhalo 476171 has a three-dimensional velocity dispersion of 
280 km s$^{-1}$ and a black hole of mass $\approx 2 \times 10^9\, M_\sun$. 
These values are only a little smaller 
than those that have been used to define massive ultracompact galaxies
\citep[MUGs,][]{Buitragoetal2018}. To establish whether 476171 can actually be 
classified as a MUG is beyond the scope of this paper. It is also of interest to
measure the stellar and dynamical masses of MUGs to determine whether 
some or all them are dark matter deficient galaxies.
      
\begin{figure}
	\centering
	\includegraphics[width=0.32\linewidth]{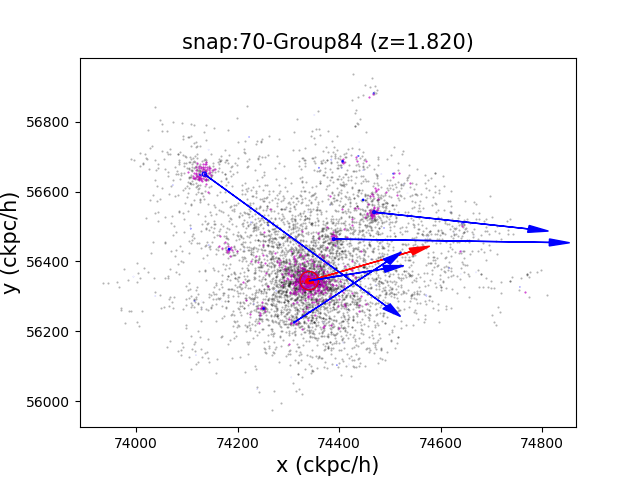}
	\includegraphics[width=0.32\linewidth]{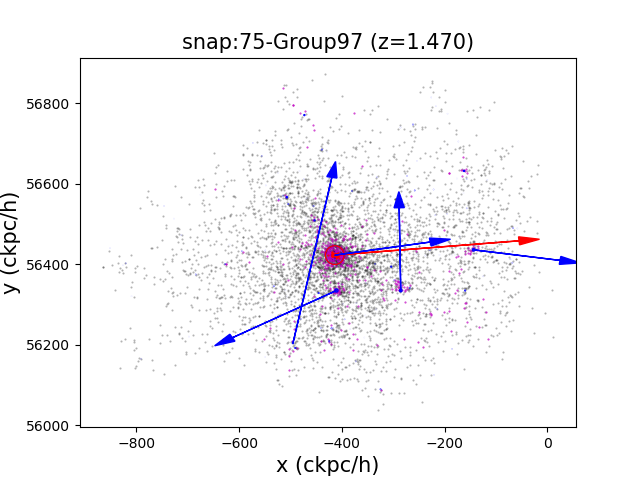}
	\includegraphics[width=0.32\linewidth]{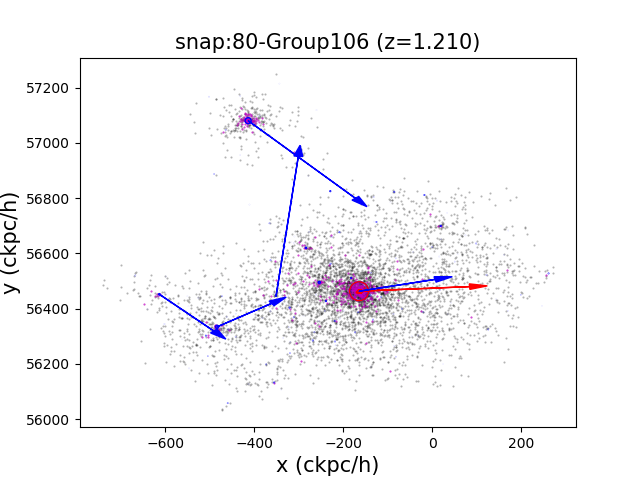}\\
	
	\includegraphics[width=0.32\linewidth]{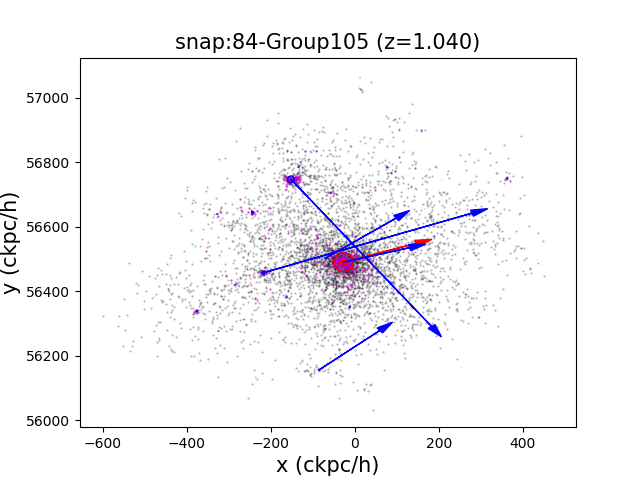}
	\includegraphics[width=0.32\linewidth]{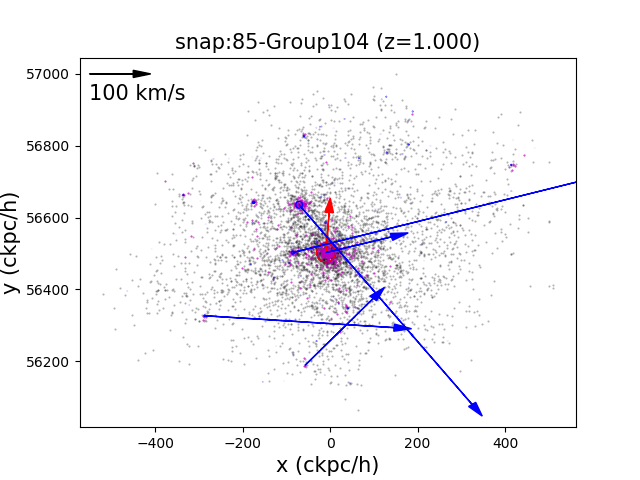}
	\includegraphics[width=0.32\linewidth]{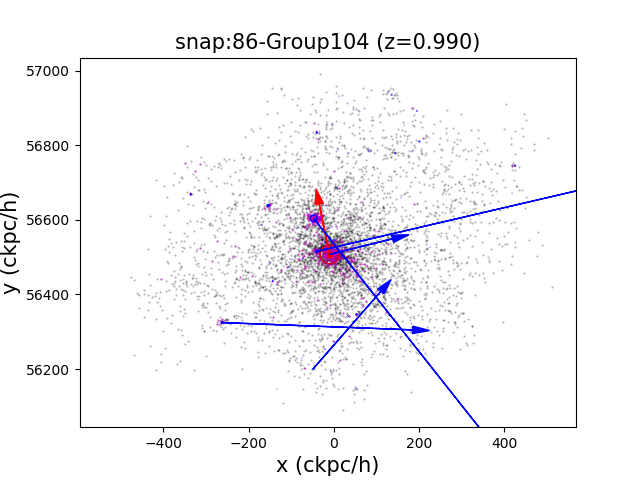}\\
	
	\includegraphics[width=0.32\linewidth]{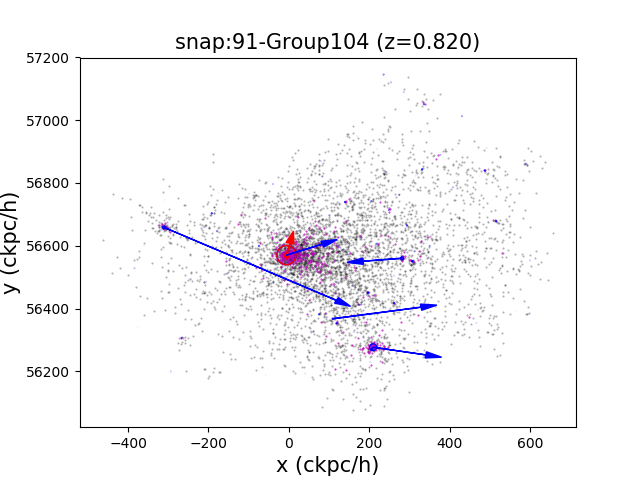}
	\includegraphics[width=0.32\linewidth]{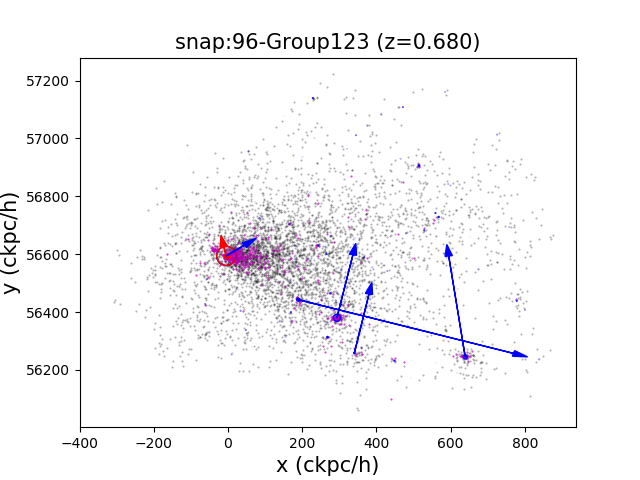}
	\includegraphics[width=0.32\linewidth]{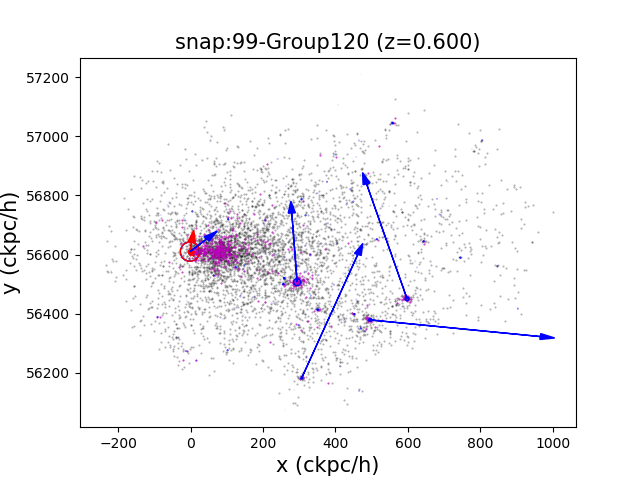}\\
	
	\includegraphics[width=0.32\linewidth]{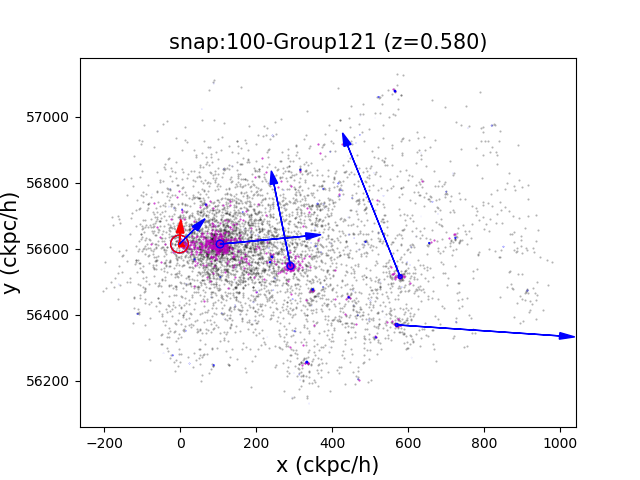}
	\includegraphics[width=0.32\linewidth]{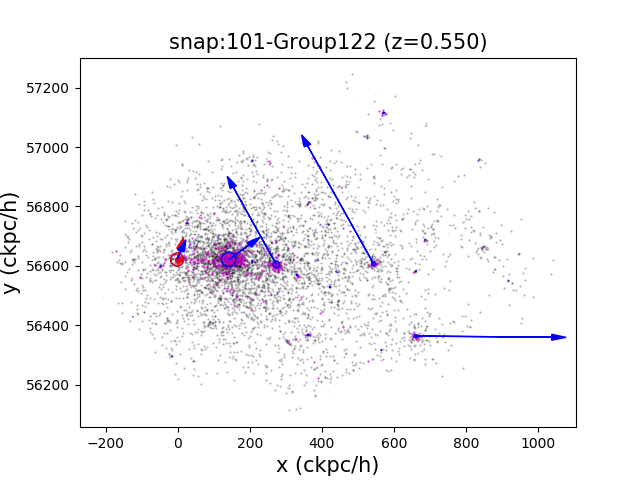}
	\includegraphics[width=0.32\linewidth]{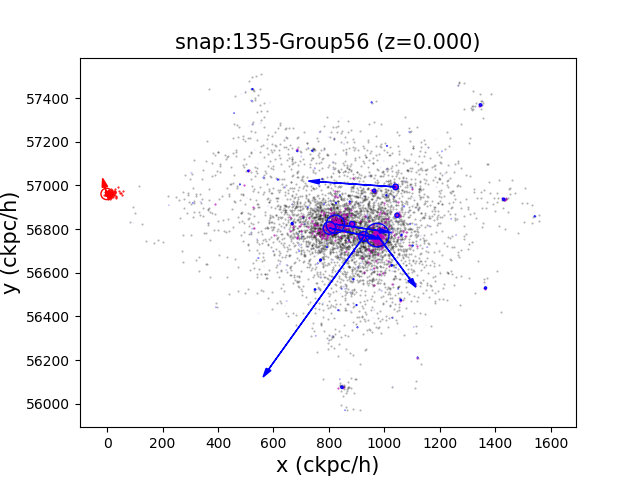}\\
	\caption{The evolutionary history of subhalo 476171. See text for information about symbols. Distance units are comoving kpc$/h$ where $h = 0.704$. The arrow in the top-left corner of the central panel in the second row indicates the velocity scale.}
	\label{fig:cutout7084}
\end{figure}

To understand how galaxy subhalo 476171 formed, we trace its formation 
history by examining the cutouts (which give the matter distribution and 
properties, such as the positions and velocities of the gas, dark matter 
and star particles, of its progenitors and parent groups) in snap 54 to 103 
(from $z = 4.01$ to 0.5) and in snap 134 ($z=0.01$) and 135 ($z = 0$). 
Figure \ref{fig:cutout7084} plots the cutout data of 476171 and its parent 
groups in a dozen snaps. The gray dots represent dark matter particles, 
purple dots stars, blue circles subhalos 
(with circle area proportional to stellar mass), and blue arrows represent 
the velocities of the five most massive subhalos in the parent groups. The red 
circles identify the progenitors of 476171 in the different snaps, the red 
dots are the positions of the stars which finally end up in subhalo 476171 
in snap 135, and red arrows represent the mean velocity of the red stars. 
From these figures we see that at first the 
subhalo is at the center of its parent group that has a very large dark 
matter halo and they largely move together in the positive $x$ direction. But 
in snap 85 ($z=1.0$), the subhalo has ceased moving in the positive $x$ 
direction while the rest of the group largely continues to do so. The subhalo 
then moves slowly towards the edge of the dark matter halo although it still 
dominates the whole halo until snap 100 ($z=0.58$) when a new subhalo starts 
forming at the center of the dark matter halo. This new subhalo comes to 
dominate the rest of the dark matter while the progenitor subhalo of 476171 
is now at the edge of the whole dark matter halo and continues to move 
outwards. By this point the progenitor subhalo of 476171 has lost most of its 
dark matter and some (although relatively less) of its stellar mass which 
results in it becoming a dark matter deficient galaxy subhalo. (See the 
lower redshift parts of the central and lower subpanels in the subhalo 
476171 panel of Fig.\ \ref{fig:mhs}.) It keeps moving outward, 
getting farther 
and farther away from the rest of the dark matter halo. By snap 135
at $z = 0$, it has completely escaped from its former parent group, forming
an isolated group of one with much more stellar mass than dark matter mass.

\begin{figure}
	\includegraphics[width=1.3\textwidth,angle=90]{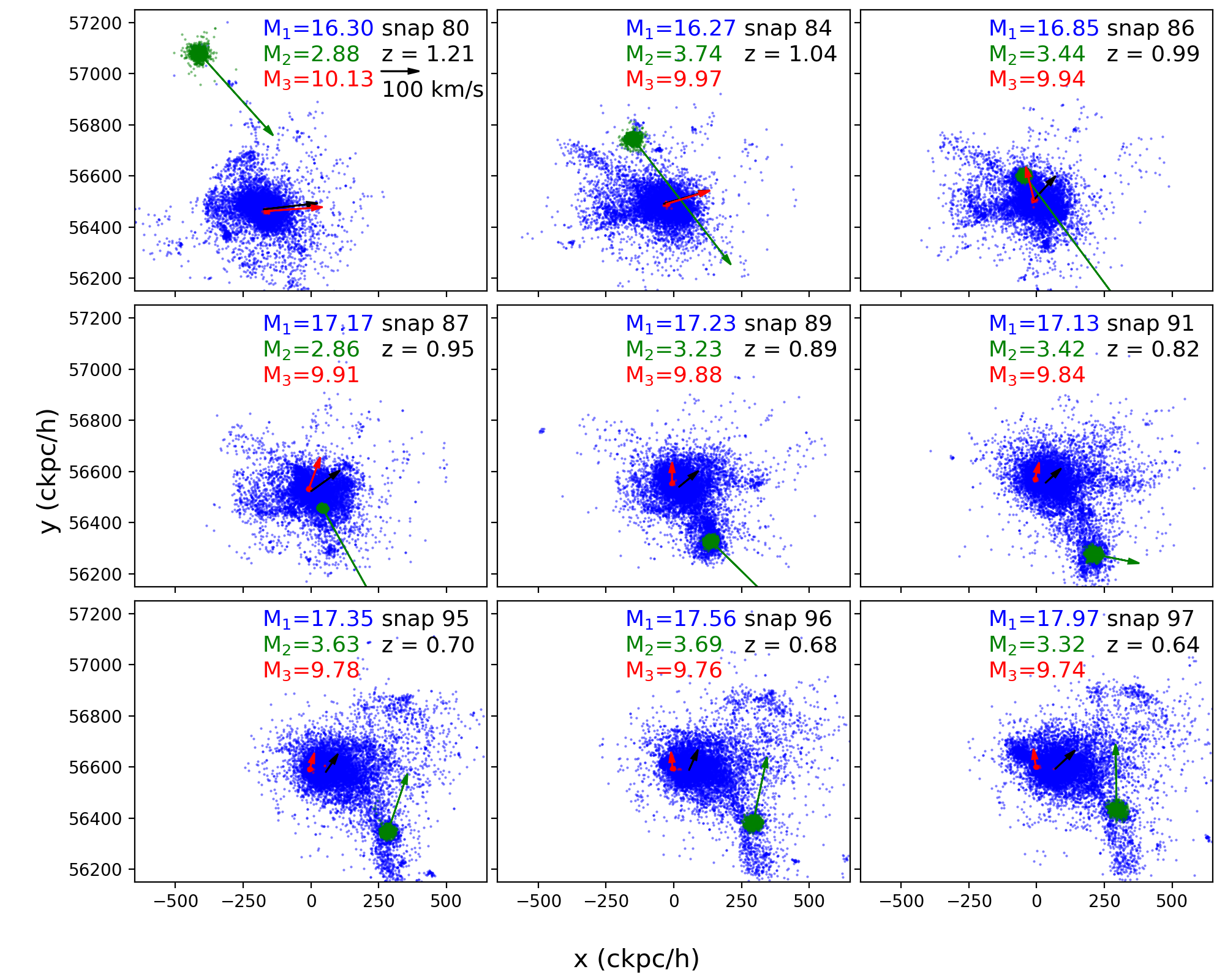}\\
	\caption{The stellar evolutionary history of subhalo 476171 and it progenitors, from snap 80 to 97. See text for information about symbols. Distance units are comoving kpc$/h$ where $h = 0.704$. }\label{fig:merger1}
\end{figure}
\begin{figure}
	\includegraphics[width=1.3\textwidth,angle=90]{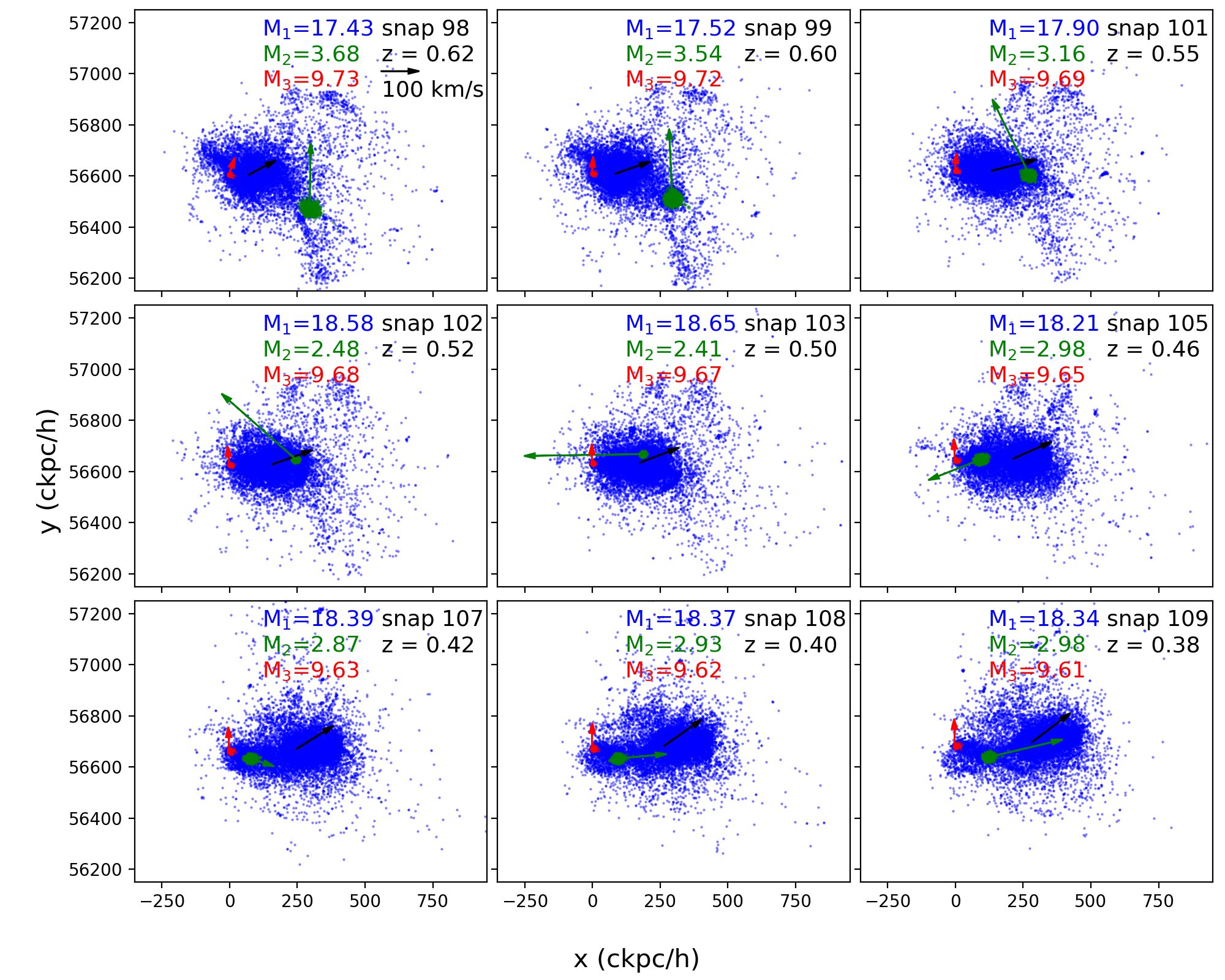}\\
	\caption{Same as Fig.\ \ref{fig:merger1} but now from snap 98 to 109.}\label{fig:merger2}
\end{figure}

It is important to more clearly understand the physical processes that caused 
the ejection of galaxy subhalo 476171, as well as those that 
played a role in causing it to become dark matter deficient. We do not yet 
have a satisfactory understanding of either of these. In an attempt to 
elucidate what took place we focus on the progenitor(s) of galaxy subhalo 
476171 and the
second most massive nearby galaxy subhalo, whose SUBFIND ID is 78979 in snap 80
($z = 1.21$), and follow these to snap 110 ($z = 0.38$). Eighteen snaps between 
80 and 110 are shown in Figs.\ \ref{fig:merger1} and \ref{fig:merger2}. For
clarity these figures show only the more centrally concentrated stellar 
particles. In these two groups of 
figures, the green dots represent the stellar particles of subhalo 78979 
and its descendants, the red dots are the stellar particles which will leave 
the FoF group and eventually end up in the subhalo 476171, and the blue dots 
(along with the red dots) are the stellar particles in the progenitors of 
galaxy subhalo 476171. (At snap 99 the blue and red dots belong to subhalo 
143860, after snap 100 most of the blue dots belong to another subhalo, 
subhalo 155159 at snap 101.) The green, red, and black arrows are the 
velocities of the green, red, and blue collections of stars. The arrow in 
the top-left panels of Figs.\ \ref{fig:merger1} and \ref{fig:merger2} shows the
velocity scale. In each panel we also list the snap number, the 
redshift, and the stellar mass (in units of $10^{10} M_\sun$) of each of 
the three star 
clumps (in three different colors). Table \ref{tab:mass} lists these and other 
relevant masses.\footnote{In Figs.\ \ref{fig:merger1} and \ref{fig:merger2} and 
Table \ref{tab:mass} the listed masses are determined by summing together the 
masses of all the particles of interest. These are not SubhaloMassType masses, 
although they do not appreciably differ from SubhaloMassType masses when both 
can be computed. We note that unlike SubhaloMassType masses, where gas masses
include wind particle masses, in our procedure wind particle masses are 
included with star masses. (Wind particle masses do not contribute 
significantly to the totals.) We need to use our procedure for the mass 
determination (instead of using SubhaloMassType) because we want to follow the
red stellar mass that eventually ends up in subhalo 476171 from well before the 
time when subhalo 476171 came into existence.}     

From these figures we see that the green and red+blue subhalos participate 
in a significant and protracted interaction.\footnote{By also viewing the 
$x$-$z$ projections of snaps 80 to 110 we have verified that the green 
subhalo passes through the red+blue subhalo.}
The first part of this interaction, during snaps 84 to 89, results in the 
braking of the positive $x$ direction motion of the red clump relative to 
the blue stellar particles. Given the length scales involved it is reasonable
to assume that the gravitational force is the main cause of the relative motion
of the green and red+blue stars and also of the relative motion between the 
red clump and the blue stars. See Table \ref{tab:mass} for the masses involved 
in these motions. It seems that as the green clump comes in to the red+blue halo
(see snaps 84 and 86) it initially more successfully pulls the compact red 
clump to it, compared to the effect it has on the more diffuse blue star 
collection (which it distorts by more effectively pulling closer blue stars
towards it), and this gives rise to an overall relative velocity between the 
red and blue star particles. 

By snap 95 ($z = 0.70$), the green subhalo has almost turned around and as it 
moves back farther into the red+blue subhalo it has a bigger effect on the 
nearer blue particles (the red clump is now farther away), pulling them towards 
it and causing 
the blue particles to accelerate in the positive $x$ direction and away from 
the red clump. Until about snap 103 ($z = 0.5$) the green subhalo is on 
the right of, and closer to, the center of the blue stars and so 
continues to more effectively pull blue stars away from the red clump. The 
green 
subhalo continues to move in a counterclockwise direction and during snaps 
104 to 109 it is between the red clump and the center of the blue stars 
and no longer seems to increase the relative velocity between the blue and red 
clumps, however by this time the relative velocity is large enough for the 
red clump to 
go its own way and escape from the blue clump.

\begin{figure}
\includegraphics[width=1.0\textwidth]{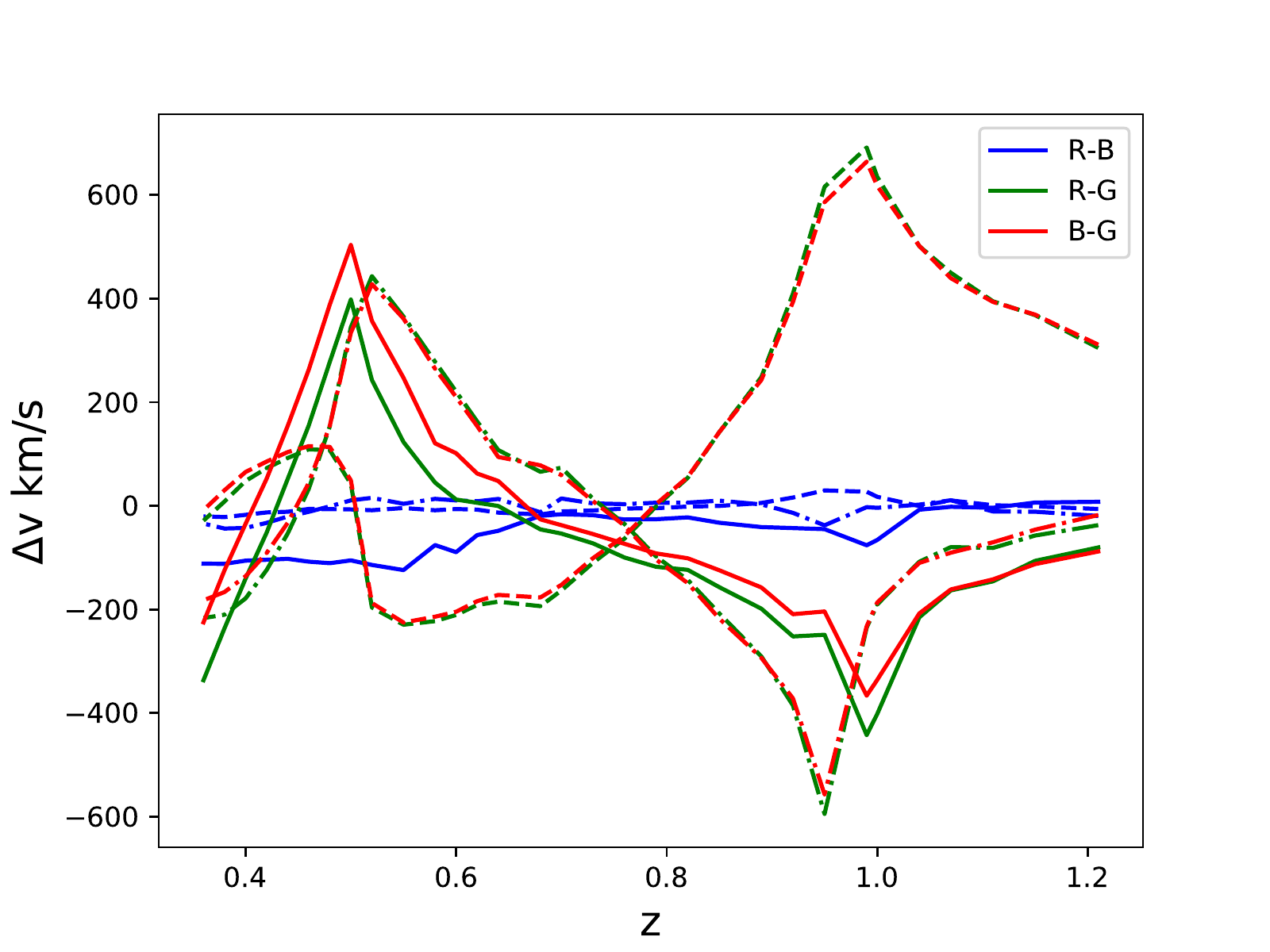}\\
\caption{Redshift evolution of the velocity differences between the red, blue,
and green collections of stars. Solid, dashed, and dash-dotted lines represent
velocity differences in the $x$, $y$, and $z$ directions. Blue lines show 
relative
velocities between red and blue clumps, green lines are relative velocities 
between red and green clumps, and red lines indicate relative velocities 
between blue and green clumps,}\label{fig:dv}
\end{figure}

This behavior is also apparent in the redshift evolution of the velocity 
differences between the red, blue, and green clumps, see 
Fig.\ \ref{fig:dv}.\footnote{We thank J.\ Peebles for suggesting that we examine
these velocity differences.}  
We see that as the green clump of stars makes a first pass through the 
central region of the red+blue
clump of stars during redshifts between about 1 and 0.9, which correspond to 
snaps 86 and 89, the $x$ component of the red and blue velocity difference
(the solid blue line in Fig.\ \ref{fig:dv}) deviates away from zero to the 
negative side, because the green clump is more effectively pulling the blue 
stars towards it. As the green clump continues to move outwards the 
differential effect it has on the red and blue clumps presumably decreases, 
resulting in a decrease of the velocity difference between the red and blue 
clumps. By snap 95 ($z = 0.7$) the green clump has turned around and starts 
moving in a counterclockwise direction. As it comes back in it is on the right
side of the blue clump and closer to it than to the red clump, presumably 
pulling more effectively on the blue clump than on the red clump and so 
increasing the $x$ component of the velocity difference between the blue and 
red clumps, that by snap 99 ($z = 0.6$) has settled down to around 100 km/s, 
with the red stars drifting away in the negative $x$ direction from the blue 
stars.

It seems reasonable, at the level of the big picture, to presume that the 
tidal effects of gravity are responsible for both the red clump being 
ejected and for it becoming dark matter deficient, and that the interaction 
between the green
and red+blue clumps was the prime cause of both, with the compactness of the 
red clump presumably also playing an important role. It is however very 
desirable to have a better and more granular understanding of this phenomenon,
which might require a local simulation with higher resolution.    

Table \ref{tab:mass} lists masses of these objects from snap 80 to 110
(we do not record black hole particle masses here as these are much smaller).
While the green clump is relatively much less massive than the red+blue one
(it is only 6\% as massive at snap 80), they have a reasonably high relative 
velocity, and the green clump loses most of its dark matter while interacting
with the red+blue one. So it seems likely that gravity has enough to work with 
to be able to eject the red clump while stripping it of most 
of its dark matter and a significant amount of its stellar matter, thus
creating a massive dark matter deficient galaxy (subhalo 476171). 

Interestingly, if galaxy subhalo 476171 is a MUG, then perhaps the formation 
process we have 
summarized here can explain how some MUGs can come to exist in low-density
environments \citep{Buitragoetal2018}. Observationally determining 
the dark matter fraction of MUGs as a function of environment might allow
for discrimination between different MUG formation channels 
\citep{Buitragoetal2018}. 

\begin{table}
	\centering
	\begin{tabular}{l|c|c|c|c|c|c|c|c}
		\hline 
		\multirow{2}*{Snap (Redshift)} & \multicolumn{3}{c|}{Red+Blue} & \multirow{2}*{Blue SW} & \multirow{2}*{Red SW} & \multicolumn{3}{c}{Green} \\
		\cline{2-4} \cline{7-9}
		& Gas & DM & SW	& & & Gas & DM & SW	\\ \hline
80 (1.21) &	17.2	&	876	&	26.4	&	16.3	&	10.1	&	9.81	&	42.5	&	2.88	 \\ \hline
81 (1.15) &	18.7	&	892	&	26.3	&	16.2	&	10.1	&	9.49	&	37.5	&	3.22	 \\ \hline
82 (1.11) &	19.8	&	901	&	26.3	&	16.2	&	10.1	&	9.05	&	35.5	&	3.33	 \\ \hline
83 (1.07) &	22.5	&	929	&	26.2	&	16.2	&	10	&	7.18	&	26.2	&	3.51	 \\ \hline
84 (1.04) &	25.2	&	942	&	26.27	&	16.3	&	9.97	&	6.05	&	23.2	&	3.74	 \\ \hline
85 (1) &	29.6	&	975	&	26.34	&	16.4	&	9.94	&	3.97	&	15.8	&	3.67	 \\ \hline
86 (0.99) &	31.6	&	988	&	26.74	&	16.8	&	9.94	&	2.9	&	11.3	&	3.44	 \\ \hline
87 (0.95) &	35	&	1010	&	27.11	&	17.2	&	9.91	&	1.87	&	5.69	&	2.86	 \\ \hline
88 (0.92) &	34	&	1010	&	27	&	17.1	&	9.9	&	3.31	&	8.13	&	3.17	 \\ \hline
89 (0.89) &	37.1	&	1040	&	27.08	&	17.2	&	9.88	&	2.23	&	7.64	&	3.23	 \\ \hline
90 (0.85) &	38.7	&	1070	&	27.06	&	17.2	&	9.86	&	3.11	&	8.17	&	3.37	 \\ \hline
91 (0.82) &	41.5	&	1100	&	26.94	&	17.1	&	9.84	&	1.98	&	8.39	&	3.42	 \\ \hline
92 (0.79) &	43	&	1110	&	26.93	&	17.1	&	9.83	&	2.45	&	8.73	&	3.49	 \\ \hline
93 (0.76) &	23.4	&	1120	&	26.91	&	17.1	&	9.81	&	2.67	&	9.26	&	3.51	 \\ \hline
94 (0.73) &	11.3	&	1130	&	27.09	&	17.3	&	9.79	&	3.24	&	8.94	&	3.51	 \\ \hline
95 (0.7) &	11.5	&	1120	&	27.18	&	17.4	&	9.78	&	3.57	&	8.98	&	3.63	 \\ \hline
96 (0.68) &	12.4	&	1130	&	27.36	&	17.6	&	9.76	&	2.1	&	10.1	&	3.69	 \\ \hline
97 (0.64) &	13.8	&	1150	&	27.74	&	18	&	9.74	&	1.06	&	8.23	&	3.32	 \\ \hline
98 (0.62) &	14.7	&	1160	&	27.13	&	17.4	&	9.73	&	1.29	&	10.7	&	3.68	 \\ \hline
99 (0.6) &	14.8	&	1160	&	27.22	&	17.5	&	9.72	&	0.935	&	9.41	&	3.54	 \\ \hline
100 (0.58) &	14.9	&	1110	&	23.11	&	13.4	&	9.71	&	0.781	&	7.85	&	3.32	 \\ \hline
101 (0.55) &	15.9	&	1160	&	27.59	&	17.9	&	9.69	&	0.81	&	6.49	&	3.16	 \\ \hline
102 (0.52) &	16.7	&	1190	&	28.28	&	18.6	&	9.68	&	0.601	&	3.6	&	2.48	 \\ \hline
103 (0.5) &	16.8	&	1190	&	28.37	&	18.7	&	9.67	&	0.737	&	3.28	&	2.41	 \\ \hline
104 (0.48) &	17.5	&	1220	&	27.86	&	18.2	&	9.66	&	0.606	&	4.99	&	2.92	 \\ \hline
105 (0.46) &	17.3	&	1240	&	27.85	&	18.2	&	9.65	&	0.48	&	5.25	&	2.98	 \\ \hline
106 (0.44) &	18.9	&	1270	&	27.94	&	18.3	&	9.64	&	0.458	&	4.95	&	2.89	 \\ \hline
107 (0.42) &	20.5	&	1310	&	28.03	&	18.4	&	9.63	&	0.51	&	4.88	&	2.87	 \\ \hline
108 (0.4) &	19.1	&	1330	&	28.02	&	18.4	&	9.62	&	0.56	&	5.13	&	2.93	 \\ \hline
109 (0.38) &	20.5	&	1370	&	27.91	&	18.3	&	9.61	&	0.563	&	5.34	&	2.98	 \\ \hline
110 (0.36) &	22.1	&	1400	&	27.9	&	18.3	&	9.6	&	0.575	&	5.46	&	3.03	 \\ \hline

	\end{tabular} 
	\caption{Gas, dark matter (DM), and star + wind (SW) masses (in units of $10^{10} M_\sun$) of the red+blue and green clumps, as well as the star + wind masses of the blue and red clumps from snap 80 to 110. (Stellar mas dominates the star + wind masses.) The ``Red+Blue" columns give the masses of different particles in the subhalo (subhalos after snap 100) which the red and blue stellar particles belong to and the ``Green" columns give the masses of different particles in the subhalo which the green stars belong to. The ``Blue SW" and ``Red SW" columns are the sum of masses of the blue dots and red dots in Figs.\ \ref{fig:merger1} and \ref{fig:merger2}.}\label{tab:mass}
\end{table}

\section{Conclusion}

Motivated by the findings of \citet{vanDokkumetal2018a}, we have searched
for dark matter deficient galaxy subhalos in the Illustris 
simulation \citep[www.illustris-project.org,][]{Vogelsbergeretal2014a}  
of the $\Lambda$CDM cosmogony and have discovered a significant number 
that seem to have properties similar to those ascribed to ultra-diffuse 
galaxy NGC1052-DF2 by \citet{vanDokkumetal2018a}. It is of interest to more 
carefully examine such galaxy subhalos in the Illustris (and future) 
simulation(s).\footnote{Perhaps a detailed analysis of the Illustris raw 
particle data can provide more information about the dark matter deficient 
galaxy subhalos we have found, especially those subhalos which first 
appear at snap 135.} 
It would be useful to better characterize them, to understand how they form, 
and to have images of them. On the observational side, if the 
$\Lambda$CDM model and the Illustris simulation are reasonably accurate (and 
there is no evidence to suggest otherwise), it is of great interest 
to find such dark matter deficient galaxies (for example, more than 0.7\% of  
SubhaloMassType $M_{\rm star} \approx 2 \times 10^8\, M_\sun$ galaxies should 
have SubhaloMassType $M_{\rm dm}/M_{\rm star}<1$); they should provide an unusual 
and valuable perspective on cosmological structure formation and dark matter.

The dark matter deficient galaxy subhalo 476171 is much more massive and an 
outlier, but the Illustris simulation of the $\Lambda$CDM model results in
many dark matter deficient ($M_{\rm dm}/M_{\rm star}<1$) massive 
($M_{\rm star} > 10^{10}\, M_\sun$) galaxy subhalos. By studying the evolutionary
history of subhalo 476171, we see that its progenitor massive subhalo 
contained a tightly bound cluster of stars. For some reason, perhaps as a 
consequence of gravitational interaction, this tightly-bound star cluster is 
kicked out of and escapes from its parent group. Perhaps because some of these 
stars are so tightly bound together, the massive subhalo 476171 and its 
stellar content largely preserve their structural integrity during the 
ejection process, losing only some stars but almost all dark matter, 
and so resulting in a dark matter deficient massive galaxy subhalo.

While it is clearly important to understand how the dark matter deficient 
massive galaxy subhalo 476171 (and its less extreme cousins) formed, if 
the Illustris simulation and the $\Lambda$CDM structure formation model 
are accurate then such objects should exist in the 
real universe (for example, more than 0.4\% of galaxies with SubhaloMassType  
$M_{\rm star} \approx 10^9\, M_\sun$ should have SubhaloMassType
$M_{\rm dm}/M_{\rm star}<1$). Due to the large stellar mass of subhalo 476171, 
its r band magnitude is about $-21$, bright enough for us to detect it. It is
of great interest to search for such objects; perhaps these are the MUGs.

We are indebted to F.\ Marinacci, J.\ Peebles, V.\ Springel, and 
M.\ Vogelsberger for valuable advice, comments, and discussion. We 
acknowledge valuable discussions with P.\ Gagrani, C.-G.\ Park, 
L.\ Samushia, and L.\ Weaver. This work is supported by the National 
Basic Research Program of China (973 Program, grant No. 2014CB845800), 
the National Natural Science Foundation of China (grants 11422325, 
11373022, and U1831207), the Excellent Youth Foundation of Jiangsu 
Province (BK20140016), and by DOE grant DE-SC0019038. H.Y.\ also acknowledges 
support by the China Scholarship Council for studying abroad.


\begin{thebibliography}{}
\expandafter\ifx\csname natexlab\endcsname\relax\def\natexlab#1{#1}\fi
\providecommand{\url}[1]{\href{#1}{#1}}

\bibitem[{{Alam} {et~al.}(2017)}]{Alametal2017}
{Alam}, S., {Ata}, M., {Bailey}, S., {et~al.} 2017, \mnras, 470, 2617
[arXiv:1607.03155]

\bibitem[{{Blakeslee} \& {Cantiello}(2018)}]{BlakesleeCantiello2018}
{Blakeslee}, J.~P., \& {Cantiello}, M.\ 2018, arXiv:1808.02176

\bibitem[{{Buitrago} {et~al.}(2018)}]{Buitragoetal2018}
{Buitrago}, F., {Ferreras}, I., {Kelvin}, L.~S., {et~al.} 2018, 
arXiv:1807.02534

\bibitem[{{Famaey} {et~al.}(2018)}]{Famaeyetal2018}
{Famaey}, B., McGaugh, S., \& Milgrom, M.,\ 2018, arXiv:1804.04167

\bibitem[{{Farooq} {et~al.}(2013)}]{Farooqetal2013}
{Farooq}, O., {Crandall}, S., \& {Ratra}, B.\ 2013, Phys.\ Lett.\ B, 726, 72
  [arXiv:1305.1957]

\bibitem[{{Farooq} {et~al.}(2017)}]{Farooqetal2017}
{Farooq}, O., {Madiyar}, F.~R., {Crandall}, S., \& {Ratra}, B.\ 2017, \apj,
  835, 26 [arXiv:1607.03537]

\bibitem[{{Farooq} \& {Ratra}(2013)}]{FarooqRatra2013}
{Farooq}, O., \& {Ratra}, B.\ 2013, \apj, 766, L7 [arXiv:1301.5243]

\bibitem[{{Fischler} {et~al.}(1985)}]{Fischleretal1985}
{Fischler}, W., {Ratra}, B., \& {Susskind}, L.\ 1985, Nucl.\ Phys.\ B, 259, 730

\bibitem[{{Genel} {et~al.}(2014)}]{Geneletal2014}
Genel, S., Vogelsberger, M., Springel, V., {et~al.} 2014, \mnras, 445, 175
[arXiv:1405.3749]
 
\bibitem[{{Guth} \& {Pi}(1982)}]{GuthPi1982}
{Guth}, A.~H., \& {Pi}, S.-Y.\ 1982, Phys.\ Rev.\ Lett., 49, 1110
 
\bibitem[{{Haridasu} {et~al.}(2018)}]{Haridasuetal2018}
{Haridasu}, B.~S., {Lukovi{\'c}}, V.~V., Moresco, M., \& {Vittorio}, N.\ 2018,
arXiv:1805.03595

\bibitem[{{Hawking}(1982)}]{Hawking1982}
{Hawking}, S.~W.\ 1982, Phys.\ Lett.\ B, 115, 295

\bibitem[{{Hinshaw} {et~al.}(2013)}]{Hinshawetal2013}
{Hinshaw}, G., {Larson}, D., {Komatsu}, E., {et~al.} 2013, \apjs, 208, 19 
[arXiv:1212.5226]

\bibitem[{{Jesus} {et~al.}(2018)}]{Jesusetal2018}
{Jesus}, J.~F., {Holanda}, R.~F.~L., \& {Pereira}, S.~H.\ 2018, JCAP, 1805, 073
[arXiv:1712.01075]

\bibitem[{{Laporte} {et~al.}(2018)}]{Laporteetal2018}
{Laporte}, C.~F.~P., Agnello, A., \& Navarro, J.~F.,\ 2018, arXiv:1804.04139

\bibitem[{{Lukovi{\'c}} {et~al.}(2018)}]{Lukovicetal2018}
{Lukovi{\'c}}, V.~V., {Haridasu}, B.~S., \& {Vittorio}, N.\ 2018, 
arXiv:1801.05765

\bibitem[{{Martin}(2012)}]{Martin2012}
Martin, J.\ 2012, C.\ R.\ Physique, 13, 566 [arXiv:1205.3365]

\bibitem[{{Martin} {et~al.}(2018)}]{Martinetal2018}
{Martin}, N.~F., Collins, M.~L.~M,, Longeard, N., \& Tollerud, E.\ 2018, 
\apj, in press [arXiv:1804.04136]

\bibitem[{{Mitra} {et~al.}(2018)}]{Mitraetal2018}
{Mitra}, S., {Choudhury}, T.~R., \& {Ratra}, B.\ 2018, \mnras, 479, 4566 
[arXiv:1712.00018]

\bibitem[{{Moresco} {et~al.}(2016)}]{Morescoetal2016}
{Moresco}, M., {Pozzetti}, L., {Cimatti}, A., {et~al.} 2016, JCAP, 1605, 014
[arXiv:1601:01701]

\bibitem[{{Nusser} (2018)}]{Nusser2018}
{Nusser}, A.\ 2018, arXiv:1806.01812

\bibitem[{{Ooba} {et~al.}(2018a)}]{Oobaetal2018a}
{Ooba}, J., {Ratra}, B., \& {Sugiyama}, N.\ 2018a, \apj, 864, 80 
[arXiv:1707.03452]

\bibitem[{{Ooba} {et~al.}(2017)}]{Oobaetal2017}
{Ooba}, J., {Ratra}, B., \& {Sugiyama}, N.\ 2017, arXiv:1710.03271

\bibitem[{{Ooba} {et~al.}(2018b)}]{Oobaetal2018b}
{Ooba}, J., {Ratra}, B., \& {Sugiyama}, N.\ 2018b, \apj, in press
[arXiv:1712.08617]

\bibitem[{{Ooba} {et~al.}(2018c)}]{Oobaetal2018c}
{Ooba}, J., {Ratra}, B., \& {Sugiyama}, N.\ 2018c, arXiv:1802.05571

\bibitem[{{Park} \& {Ratra}(2018a)}]{ParkRatra2018a}
{Park}, C.-G., \& {Ratra}, B.\ 2018a, arXiv:1801.00213

\bibitem[{{Park} \& {Ratra}(2018b)}]{ParkRatra2018b}
{Park}, C.-G., \& {Ratra}, B.\ 2018b, arXiv:1803.05522

\bibitem[{{Park} \& {Ratra}(2018c)}]{ParkRatra2018c}
{Park}, C.-G., \& {Ratra}, B.\ 2018c, arXiv:1807.07421

\bibitem[{{Park} \& {Ratra}(2018d)}]{ParkRatra2018d}
{Park}, C.-G., \& {Ratra}, B.\ 2018d, arXiv:1809.03598

\bibitem[{{Peebles}(1982)}]{Peebles1982}
{Peebles}, P.~J.~E.\ 1982, \apj, 263, L1

\bibitem[{{Peebles}(1984)}]{Peebles1984}
{Peebles}, P.~J.~E.\ 1984, \apj, 284, 439

\bibitem[{{Planck Collaboration}(2016)}]{PlanckCollaboration2016}
{Planck Collaboration}, {Ade}, P.~A.~R., {Aghanim}, N., {Arnaud}, M.,
{et~al.} 2016, \aap, 594, A13 [arXiv:1502.01589]

\bibitem[{{Ratra} \& {Vogeley}(2008)}]{RatraVogeley2008}
{Ratra}, B., \& {Vogeley}, M.\ 2008, \pasp, 120, 235 [arXiv:0706.1565]

\bibitem[{{Ryan} {et~al.}(2018)}]{Ryanetal2018}
{Ryan}, J., {Doshi}, S., \& {Ratra}, B.\ 2018, \mnras, 480, 759 
[arXiv:1805.06408]

\bibitem[{{Scarpa} {et~al.}(2018)}]{Scarpaetal2018}
{Scarpa}, R., Hernandez, X., Martin, R.~A.~C., Falomo, R., \& 
L{\'o}pez-Corredoira, M.\ 2018, arXiv:1804.04817

\bibitem[{{Scolnic} {et~al.}(2017)}]{Scolnicetal2017}
{Scolnic}, D.~M., {Jones}, D.~O., {Rest}, A., {et~al.} 2017, 
arXiv:1710.00845

\bibitem[{{Sijacki} {et~al.}(2015)}]{Sijackietal2015}
Sijacki, D., Vogelsberger, M., Genel, S., {et~al.} 2015, \mnras, 452, 575
[arXiv:1408.6842]

\bibitem[{{Starobinsky}(1982)}]{Starobinsky1982}
{Starobinsky}, A.~A.\ 1982, Phys.\ Lett.\ B, 117, 175

\bibitem[{{Trujillo} {et~al.}(2018)}]{Trujilloetal2018}
{Trujillo}, I., {Beasley}, M.~A., {Borlaff}, A., {et~al.} 2018, 
arXiv:1806.10141

\bibitem[{{van Dokkum} {et~al.}(2018c)}]{vanDokkumetal2018c}
{van Dokkum}, P., Danieli, S., Cohen, Y., \& Conroy, C.\ 2018c, \apj, 
864, L18 [arXiv:1807.06025]

\bibitem[{{van Dokkum} {et~al.}(2018b)}]{vanDokkumetal2018b}
{van Dokkum}, P., Cohen, Y., Danieli, S., {et~al.} 2018b, \apj, 856, L30
[arXiv:1803.10240]

\bibitem[{{van Dokkum} {et~al.}(2018a)}]{vanDokkumetal2018a}
{van Dokkum}, P., Danieli, S., Cohen, Y., {et~al.} 2018a, \nat, 555, 629
[arXiv:1803.10237]

\bibitem[{{Vogelsberger} {et~al.}(2013)}]{Vogelsbergeretal2013}
Vogelsberger, M., Genel, S., Sijacki, D., {et~al.} 2013, \mnras, 436, 3031
[arXiv:1305.2913]

\bibitem[{{Vogelsberger} {et~al.}(2014a)}]{Vogelsbergeretal2014a}
Vogelsberger, M., Genel, S., Springel, V., {et~al.} 2014a, \nat, 509, 177
[arXiv:1405.1418]

\bibitem[{{Vogelsberger} {et~al.}(2014b)}]{Vogelsbergeretal2014b}
Vogelsberger, M., Genel, S., Springel, V., {et~al.} 2014b, \mnras, 444, 1518
[arXiv:1405.2921]

\bibitem[{{Wasserman} {et~al.}(2018)}]{Wassermanetal2018}
Wasserman, A., Romanowsky, A.~J., Brodie, J., {et~al.} 2018, arXiv:1807.07069

\bibitem[{{Yu} {et~al.}(2018)}]{Yuetal2018}
{Yu}, H., {Ratra}, B., \& {Wang}, F.-Y.\ 2018, \apj, 856, 3 [arXiv:1711.03437]

\end{thebibliography}
\end{document}